\documentclass[a4paper,fleqn,usenatbib]{mnras}

\usepackage{newtxtext,newtxmath}

\usepackage[T1]{fontenc}
\usepackage{ae,aecompl}

\usepackage{graphicx}	
\usepackage{amsmath}	
\usepackage{amssymb}	
\usepackage{float}
\usepackage{placeins}
\usepackage{pdflscape}
\usepackage{tabularx}

\usepackage[dvipsnames]{xcolor}
\usepackage[normalem]{ulem}
\newcommand{\mev}{MeV nuc$^{-1}$ }

\title[Modelling the accretion outburst history of Aql X-1]{A cooling neutron star crust after recurrent outbursts:\\ Modelling the accretion outburst history of Aql X-1}

\author[L. S. Ootes et al.]{
Laura S. Ootes,$^{1}$\thanks{E-mail: l.s.ootes@uva.nl (LSO)}
Rudy Wijnands,$^{1}$
Dany Page,$^{2}$
and Nathalie Degenaar$^{1}$
\\
$^{1}$Anton Pannekoek Institute for Astronomy, University of Amsterdam, Postbus 94249, 1090 GE Amsterdam, The Netherlands\\
$^{2}$Instituto de Astronom\'{i}a, Universidad Nacional Aut\'{o}noma de M\'{e}xico, M\'{e}xico, D.F. 04510, M\'{e}xico}

\date{Accepted XXX. Received YYY; in original form ZZZ}

\pubyear{2017}

\begin{document}
\label{firstpage}
\pagerange{\pageref{firstpage}--\pageref{lastpage}}
\maketitle

\begin{abstract}
With our neutron star crust cooling code {\tt NSCool} we track the thermal evolution of the neutron star in Aql X-1 over the full accretion outburst history from 1996 until 2015. For the first time, we model many outbursts (23 outbursts were detected) collectively and in great detail. This allows us to investigate the influence of previous outbursts on the internal temperature evolution and to test different neutron star crust cooling scenarios. Aql X-1 is an ideal test source for this purpose, because it shows frequent, short outbursts and thermally dominated quiescence spectra. The source goes into outburst roughly once a year for a few months. Assuming that the quiescent {\it Swift}/XRT observations of Aql X-1 can be explained within the crust cooling scenario \citep{waterhouse2016}, we find three main conclusions. Firstly, the data are well reproduced by our model if the envelope composition and shallow heating parameters are allowed to change between outbursts. This is not the case if both shallow heating parameters (strength and depth) are tied throughout all accretion episodes, supporting earlier results that the properties of the shallow heating mechanism are not constant between outbursts. Second, from our models shallow heating could not be connected to one specific spectral state during outburst. Third, and most importantly, we find that the neutron star in Aql X-1 does not have enough time between outbursts to cool down to crust-core equilibrium and that heating during one outburst influences the cooling curves of the next. 

\end{abstract}




\section{Introduction}
Many neutron stars are found in Low Mass X-ray Binaries (LMXBs). These neutron stars have relatively low magnetic fields ($B\lesssim10^{8-9}$G) and accrete matter from their low-mass companion ($M\lesssim1M_\odot$) when this companion overflows its Roche lobe. The accretion process can be persistent or episodic. In the latter case the so-called transient systems alternate between phases of accretion (outbursts) and quiescence. Neutron stars in X-ray transients are of particular interest, because in the quiescence phase several properties of neutron stars can be studied. 

During outburst, the crust of the neutron star is heated up due to various accretion-induced processes such as electron captures, beta-decay, and pycnonuclear reactions \citep[e.g.][]{sato1979,haensel1990}. These deep crustal heating processes release in total $1.5-2.0$ MeV per accreted nucleon in the inner crust of the neutron star \citep{haensel1990,haensel2003,haensel2008}. Additionally, crust cooling studies have shown that there must be an additional shallow heat source that releases typically $1-2$ \mev  in the outermost layers of the crust \citep[e.g.][]{brown2009}. Even though heat can flow from the crust into the core, the core is heated only marginally on the time scales of outburst durations because of its high heat capacity \citep{colpi2001,wijnands2013}. Consequently, the crust can become hotter than the core during outburst and hence the two become thermally decoupled. Once accretion halts, the crust will cool down to restore crust-core equilibrium \citep{brown1998,rutledge2002b}. 

To date, the expected crust cooling has been observed from seven systems \citep[e.g.][]{wijnands2002,wijnands2004,fridriksson2010,homan2014,degenaar2011a,degenaar2011b,degenaar2015}, potentially nine \citep{waterhouse2016,parikh2017}; see \citet{wijnands2017} for a review. These sources all showed a decrease in the amount of blackbody radiation emitted after the end of their accretion outburst which implies a decreasing surface temperature. Observational campaigns covering several years after the end of the outbursts allowed for construction of cooling curves for each of the sources, probing the temperatures of increasingly deeper layers of the crust as the time runs \citep{brown2009}. How fast the crust cools down during quiescence depends on the amount of heat stored in the crust during the outburst, on the core temperature, and on crustal properties such as composition, thermal conductivity, and crust thickness. Comparing observations of crust cooling with physical models is therefore a powerful tool to constrain crustal properties \citep{rutledge2002b,brown2009,page2013}.

Over the years multiple models have been developed that can track the thermal evolution of accreting neutron stars \citep[e.g.][]{colpi2001,shternin2007,brown2009,page2013,turlione2015}. Such codes take into account different heating processes during outburst, as well as cooling processes in outburst and quiescence via photon and neutrino emission. Adjusting the physical conditions of the neutron star in these codes in order to match the observations of cooling crusts revealed some unexpected results. First of all, observations showed that neutron stars cool down rapidly in the first few hundred days after the end of the outburst. This requires large amounts of heat to be transported inwards in a short time, which implies that accreting neutron stars have crusts with high conductivity \citep{wijnands2004,brown2009}. This required adjustment from the initial theory that the accretion of matter onto the crust would lead to a lattice structure with a significant amount of impurities and therefore a low conductivity \citep{schatz1999,brown2000}. Additionally, several sources were observed to be much hotter than theoretically expected during the crust cooling phase, specifically during the first $\sim100$ days hereof. They showed such a great difference in initial crust temperature compared to the base level (i.e. the observed temperature when crust-core equilibrium is reached in quiescence) that it has become clear that there must be an extra heating source active -- besides deep crustal heating--  during accretion in these sources that deposits heat at shallow depths \citep[e.g.][]{brown2009}. The physical origin of the shallow heating is unknown, but it must release about $1-2$ \mev  in most sources that need shallow heat to explain the observations \citep[e.g.][]{brown2009,degenaar2011b,page2013,merritt2016}. An exception is MAXI J0556-332, which requires $6-17$ \mev \citep{deibel2015, parikh2018}.

The first four sources in which crust cooling was observed were quasi-persistent sources with outburst durations $>1$ year. Such sources, in contrast to ordinary transients which have outbursts lasting $\sim$ months, were prime candidates to detect this phenomenon. This is because for the same accretion rate more heat can be stored in the crust in the case of longer outburst durations, leading to more significant cooling curves. However, the fifth source in which crust cooling was detected is IGR J17480-2446 \citep{degenaar2011,degenaar2011b,degenaar2013}, after an outburst that lasted only $\sim7$ weeks. This proved that also during the more common outbursts that last a few weeks, enough heat can be deposited in the crust to have an observable effect. 

\subsection{Aql X-1 and its potential crust cooling}
Recently, \citet{waterhouse2016} argued that the observations of Aql X-1 taken in quiescence can also be interpreted as crust cooling. Aql X-1 is a well-studied transient neutron star LMXB with regular outburst durations and relatively short recurrence times. The earliest observations of the source stem from 1965 \citep{friedman1967} and ever since many outbursts have been observed with different X-ray telescopes \citep[see e.g.][and references therein]{kaluzienski1977, campana2013}. In the past two decades the {\it Rossi X-ray Timing Explorer} (\textit{RXTE}), Monitor of All-Sky X-ray Image (MAXI), and \textit{Neil Gehrels Swift Observatory} (\textit{Swift}) have collectively provided well sampled monitoring of Aql X-1's behaviour. The source is located at a distance of $5$ kpc \citep{rutledge2001,jonker2004} and goes into outburst for $\sim70$ days roughly once a year. Note that these are average values, the actual outburst duration and recurrence time can be significantly different (see Table \ref{tab:outburstproperties}). Thermonuclear type-I X-ray bursts detected during outburst identify the accretor in this system as a neutron star \citep{koyama1981}. 

The origin of the quiescent emission of Aql X-1 is an ongoing debate \citep[see e.g.][]{rutledge2002a,brown2002,campana2003,cackett2011,cotizelati2014,waterhouse2016}. Both short ($<10^4$ s) and long-term (months-years, sometimes with outbursts in between different quiescent observations) variability has been observed in the source. Suggested causes hereof include variable low-level accretion, difference in envelope compositions (assuming that there is no low-level accretion in quiescence, this interpretation only applies to long scale variability, i.e. the envelope composition can be different after every new outbursts), and thermal relaxation of the crust. Thermal relaxation of the crust has often been excluded as origin of  decreases in luminosity for Aql X-1 for two reasons. First of all, \citet{brown1998} suggested that because of the low outburst fluence (integrated outburst luminosity), the crust cannot be heated significantly during outburst by deep crustal heating. Second, \citet{ushomirsky2001} found that quiescent variability can only be explained as thermal relaxation if the source has enhanced neutrino emission in the core. This is inconsistent with the earlier finding that the thermal spectral component of the quiescent luminosity observed from Aql X-1 requires negligible core neutrino emissivity \citep{brown1998,colpi2001}. Therefore, \citet{ushomirsky2001} discarded the crust cooling interpretation. However, in these studies, only heating from the deep crustal heating processes was considered, while we now know that shallow heating can also significantly heat the outer layers. 

\citet{waterhouse2016} analysed observations taken with the X-Ray Telescope (XRT) on board {\it Swift} after three different outbursts and found that the spectral behaviour of the source is consistent with the predictions for crust cooling. First of all, the surface temperatures determined from the spectral fits followed a declining trend after each of the different outbursts. Second, for the epochs after two very similar outbursts (in terms of both outburst energetics and duration), very similar temperature evolution was observed, in the sense that the two potential cooling curves lined up smoothly. Finally, the third investigated outburst, which was observed to have significantly smaller outburst luminosity and a different outburst profile compared to the other two, behaved in line with the crust relaxation theory. The source was significantly cooler directly after the end of the outburst and its surface temperature decayed faster, indicating that the crust was less powerfully heated during this fainter outburst. However, despite these indications of crust cooling, the authors could not exclude that there is (also) residual accretion in quiescence, especially because short accretion flares have been observed on top of the potential cooling curve after one of the outburst \citep{cotizelati2014}. But, as \citet{waterhouse2016} noted, even if residual accretion is occurring, thermal relaxation of the crust may still dominate the quiescence evolution.  

Aql X-1 is an interesting source to test crust cooling scenarios, because of its frequent and short outbursts. Additionally, Aql X-1 alternates between hard (predominantly high energy emission) and soft states (predominantly low energy emission) during its brightest outbursts (see e.g. Figure 2 in \citet[]{waterhouse2016} and Figure 1 in \citet{ono2016}), while during fainter outbursts the source seems to reside only in the hard state. This feature of state changes might indicate a change in accretion geometry, which might influence the shallow heating processes \citep{zand2012}. \citet{waterhouse2016} modelled three of the outbursts and subsequent cooling periods and were able to place constraints on some properties such as the shallow heating strength and depth. However, the three periods were modelled individually and without taking into account  the variability in the accretion rate onto the neutron star during outburst. In our previous crust cooling research focused on KS 1731-260, we found that variations in accretion rate during outburst can strongly influence the calculated cooling curves and hence the derived crustal parameters \citep{ootes2016}. This especially affects the shallow heating strength, which is very sensitive to short time scale variations as well as to the overall decline in accretion rate near the end of the outburst. 

Here we show detailed modelling of the temperature evolution in Aql X-1 over 20 years. The continuous and frequent monitoring of the source allows us to track the full outburst history observed from 1996 until July 2015. We use our crust cooling code {\tt NSCool} \citep{page2013}, taking into account accretion rate variability \citep{ootes2016}, to model the 23 outbursts observed in this period. We model all outbursts collectively rather than individually (i.e. in one run of the code we model all 23 outbursts and quiescent episodes chronologically), to investigate the influence of the outburst history on the calculated cooling curves. We use the quiescent observations after five different outbursts to probe the crustal properties of the neutron star and investigate how these properties might change between outbursts. 

\section{Modelling Thermal Evolution}
\subsection{The {\tt NSCool} Code}
To model the thermal evolution of Aql X-1 based on its accretion history we use our crust cooling code {\tt NSCool} \citep{page2013,page2016,ootes2016}. This is a one-dimensional code (i.e. assuming spherical symmetry) that solves the energy transport and energy balance equations in a general relativistic approach. 

\subsubsection{Stellar structure}
For the stellar structure, we use one of the pre-built stellar models that is based on the Akmal-Pandharipande-Ravenhall equation of state (A18+$\delta$v+UIX*) for the core \citep{akmal1998}. This structure model assumes that the crust (defined as the density region $1.0\times10^{8}<\rho<1.5\times10^{14} \text{ g cm}^{-3}$) is composed of the burning ashes of accreted material as described by \citet{haensel2008}. In the inner crust we assume dripped neutrons to form an $^1$S$_0$ superfluid \citep{schwenk2003} with the resulting strong suppression of their specific heat. The matter at densities $\rho<10^8$ g cm$^{-1}$ is assumed to form the envelope of the neutron star. In these outermost layers, the hydrogen and helium accreted from the companion is fused into heavier elements. We take $^{56}\text{Fe}$ as final product of these nuclear reactions and consequently to form the basis of the outer crust. Fusion reactions during the accretion outburst do not contribute significantly to the heating of the crust, because the energy is liberated on the surface of the neutron star and immediately radiated away \citep[see e.g. the review by][]{chamel2008}.

\begin{figure}
   \includegraphics[width=\columnwidth]{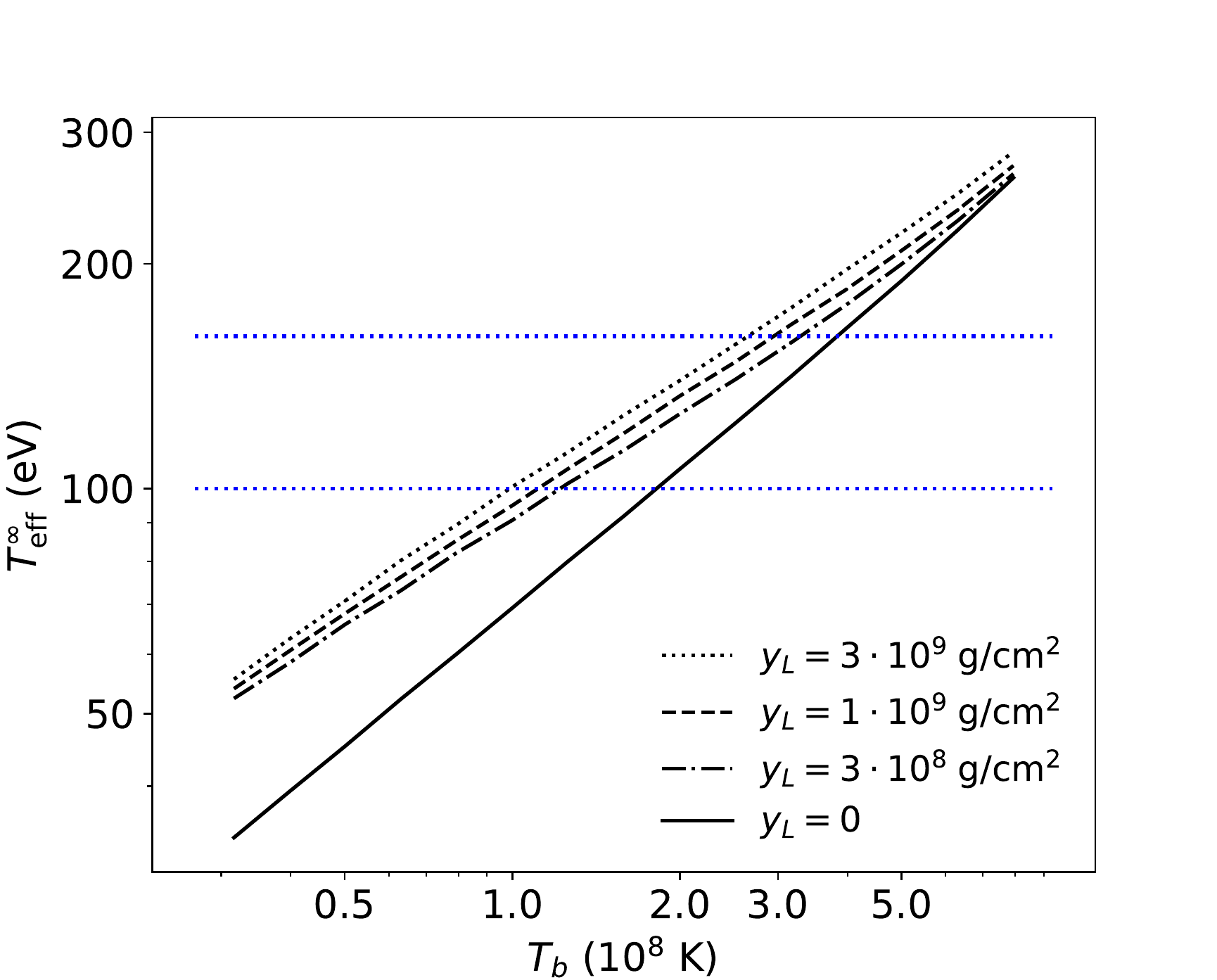}
   \caption{$T_b - T_\text{eff}^\infty$ relationships for envelopes with various amounts of light elements, indicated by their respective column densities, $y_L$, and compared to a heavy element envelope with $y_L = 0$. These models assume a stellar mass and radius of 1.6 $M_\odot$ and 11 km, respectively. The two horizontal, blue, dotted lines show the range of $T_\text{eff}^\infty$ observed during the cooling phases.}
 \label{fig:Tb-Ts}
\end{figure}

\subsubsection{The envelope}
For each time step {\tt NSCool} solves the thermal evolution equations from the centre of the star up to a boundary density $\rho_\text{b}=10^8 \text{ g cm}^{-3}$. Heat propagation trough the envelope ($\rho<\rho_\text{b}$) only depends on the structure and composition of the envelope and not on the underlying layers. Therefore, we treat the heat propagation through the envelope independently from the crust. 

The envelope composition is highly variable in time during an outburst since it depends on the accretion rate, the composition of the accreted material and the nuclear reactions taking place. The latter include hydrogen, helium, and carbon burning, which can happen either stably or unstably depending on the accretion rate \citep[see e.g. the review by][]{bildsten1998}. This leads to a shell-like envelope structure with each shell consisting of heavier elements with increasing depth. We assume that in quiescence no residual accretion takes place (and thus neglect any possible effects of accretion flares; we discuss their influence in Section \ref{discussion:flares}) and that the envelope composition at that time is thus equal to the envelope composition at the end of the preceding outburst. This is the only composition of concern for our cooling calculations, and therefore we model each full outburst and cooling curve with constant envelope composition. 

We generated envelope models in the same line as described in \citet{potekhin1997} and \citet{brown2002} but with a bottom density
$\rho_\text{b} = 10^8$ g cm$^{-3}$ (instead of $10^{10}$ g cm$^{-3}$).\footnote{We use a lower boundary density of $\rho_\text{b}=10^8\text{ g cm}^{-3}$ to be able to incorporate shallow heating into our model, which is found to be required -- for some of the other crust cooling sources --  at densities around $\sim 10^9\text{ g cm}^{-3}$.} We parametrise the layer of light elements, comprising a layer of hydrogen + helium on top of a thicker layer of $^{12}$C, by its total
column density $y_\text{L}$ (in which it is assumed that the carbon column depth is 10 times the helium column depth) and assume the presence of $^{56}$Fe in the deeper layers. These envelope models provide us with the outer boundary condition of {\tt NSCool} as a relationship between the temperature $T_\text{b}$ 
at $\rho_\text{b}$ and the redshifted effective temperature $T_\text{eff}^\infty$. The calculated $T_\text{eff}^\infty$ depends on the mass $M$ and radius $R$ of the star through the surface gravity $g_s$ as $T_\text{eff} \propto g_s^{1/4}$ (with $T\text{eff}$ the local effective temperature), and is higher if the envelope contains more light elements because their presence increases the heat conductivity compared to Fe. The resulting $T_b - T_\text{eff}^\infty$ relationships for various amounts of light elements are shown in Figure \ref{fig:Tb-Ts}.

\subsubsection{Heating and cooling}
We calculate the time-dependent accretion rate during an outburst as described in \citet{ootes2016} based on daily averaged light curve observations (details of the obtained light curve of Aql X-1 will be discussed in Section \ref{sec:lightcurve}). We assume that $1.93 \text{ MeV}$ per accreted nucleon is released in the (inner) crust due to deep crustal heating between $\rho=1.5\times10^9-3.5\times10^{13}\text{ g cm}^{-3}$ as calculated by \citet{haensel2008}. Pycnonuclear reactions contribute the major part of the energy released in the deep crustal heating process. Additionally, we allow a supplementary amount of heat, depending on the accretion rate, to be released in the outer crust to simulate the shallow heating. We model this by setting a total amount $Q_\text{sh}$ of heat to be released per accreted nucleon, resulting in a heating luminosity
\begin{equation}
H_\text{sh}(t) \equiv Q_\text{sh}  \frac{\dot M(t)}{m_\text{u}}
\label{Eq:shallow}
\end{equation}
($m_\text{u}$ being the atomic mass unit) which is distributed evenly amongst the volume elements between $\rho_\text{sh,min}$ and $\rho_\text{sh,max}=5* \rho_\text{sh,min}$. 

In quiescence, the neutron star cools down via photon emission from the surface and neutrino cooling from the core. As standard input we assume that neutrino cooling follows the `minimal cooling paradigm' \citep[i.e. no enhanced neutrino emission in the core from direct URCA processes,][]{page2004}. At the end of the outburst the neutron star has a specific temperature profile that depends on the outburst and stellar properties. As we observe the effective temperature to decrease in quiescence, the crust cools down and deeper and deeper layers of the crust can be probed \citep{brown2009}. 

However, how fast heat propagates through the crust depends on the thermal conductivity, set by the various scattering processes. We consider the contributions to the thermal conductivity of electrons scattering with: electrons \citep{shternin2006}, crystal impurities \citep{yakovlev1980}, and with phonons and ions \citep{gnedin2001}. For the contribution of the electron-impurity scattering to the conductivity we set a free impurity parameter $Q_\text{imp}$ in the crust. In the code we allow three different density regions in the crust for which independent impurity factors can be set, with transition densities defined at $\rho=4\times10^{11}\text{ g cm}^{-3}$ and $\rho=8\times10^{13}\text{ g cm}^{-3}$ such that the three layers comprise respectively the outer crust, neutron drip region, and pasta region. 

The intrinsic base level -- the boundary temperature $T_\text{b}$ at the time that crust-core equilibrium is restored -- is set by the redshifted, uniform core temperature $\tilde T_0$ prior to accretion. During an outburst most of the heat flows towards the core which will eventually increase the core temperature and thus the intrinsic base level. This is, however, a very slow process and we can easily estimates its time scale $\tau_\mathrm{th}$ as follows. With $\tilde T_0 \simeq 10^8$ K the core has a large specific heat $C_V \simeq 10^{37} - 10^{38}$ erg K$^{-1}$ \citep{wijnands2013,cumming2017}, the higher value referring to a core made of normal degenerate matter and the lower one to a superfluid core. Given the observed average mass accretion rate of $\sim 4 \times 10^{-10}$ M$_\odot$ yr$^{-1}$, deep crustal heating plus shallow heating give a long term average heating rate of at most $\langle H \rangle \sim 10^{35}$ erg s$^{-1}$. We can hence estimate \citep{wijnands2013} that $\tau_\mathrm{th} \simeq C_V T_0/ \langle H \rangle \sim$ 300 -- 3000 yrs, much longer than the 20 yrs we are exploring in this work. Notice that the observed high quiescent luminosity, $L_q \simeq 2 \times 10^{3}$ erg s$^{-1}$, implies an inefficient neutrino emission form its core (see, e.g., figure 1 in \citealt{wijnands2013}) which is, however, automatically included in the above estimate of $\tau_\mathrm{th}$. Finally, note that the observed base level of the cooling curve also depends on the envelope composition, since we do not observe $T_\text{b}$, but $T_\text{eff}^\infty$. Because of this envelope composition dependence, the observed base level can be variable on shorter timescales \citep{brown2002}.

\subsubsection{Fit parameters}
Previously, we created models with \texttt{NSCool} in which we adjusted the input parameters by hand to reduce the $\chi^2$. The code is now expanded with an automated fitting routine that reduces the total $\chi^2$ (the sum of the $\chi^2$ values of each of the cooling curves for which we have observations). This allows us to obtain more accurate results and calculate errors on each of the parameters.

In our models of Aql X-1 we fix the mass and radius according to the values used to obtain the spectral fits: $M=1.6 \text{ M}_\odot$ and $R=11$ km \citep[as used by][in their crust cooling study of this source and whose quiescence data we use in this paper; see Section \ref{qdata}]{waterhouse2016}. The free fit parameters are the initial core temperature $\tilde T_0$, the impurity factor of the crust $Q_\text{imp}$ (in different density regions), the amount ($Q_\text{sh}$) and minimal depth ($\rho_\text{sh,min}$) of the shallow heating, and the envelope light element column depth ($y_\text{L}$). We allow the light element column depth and the amount and the depth of the shallow heating to vary between different outbursts. 

The origin and properties of the shallow heating is unknown. It is generally assumed that the shallow heating is proportional with the accretion rate (see Eq. \ref{Eq:shallow}). Modelling of different crust cooling sources has shown that this heat source is not constant between the systems, and moreover is not necessarily constant between different outbursts of the same source \citep{deibel2015,parikh2018}. By allowing the amount and depth of shallow heating to vary between outbursts, we try to constrain the likelihood that this parameter is variable in time. 

\begin{figure*}
	\includegraphics[width=0.95\textwidth]{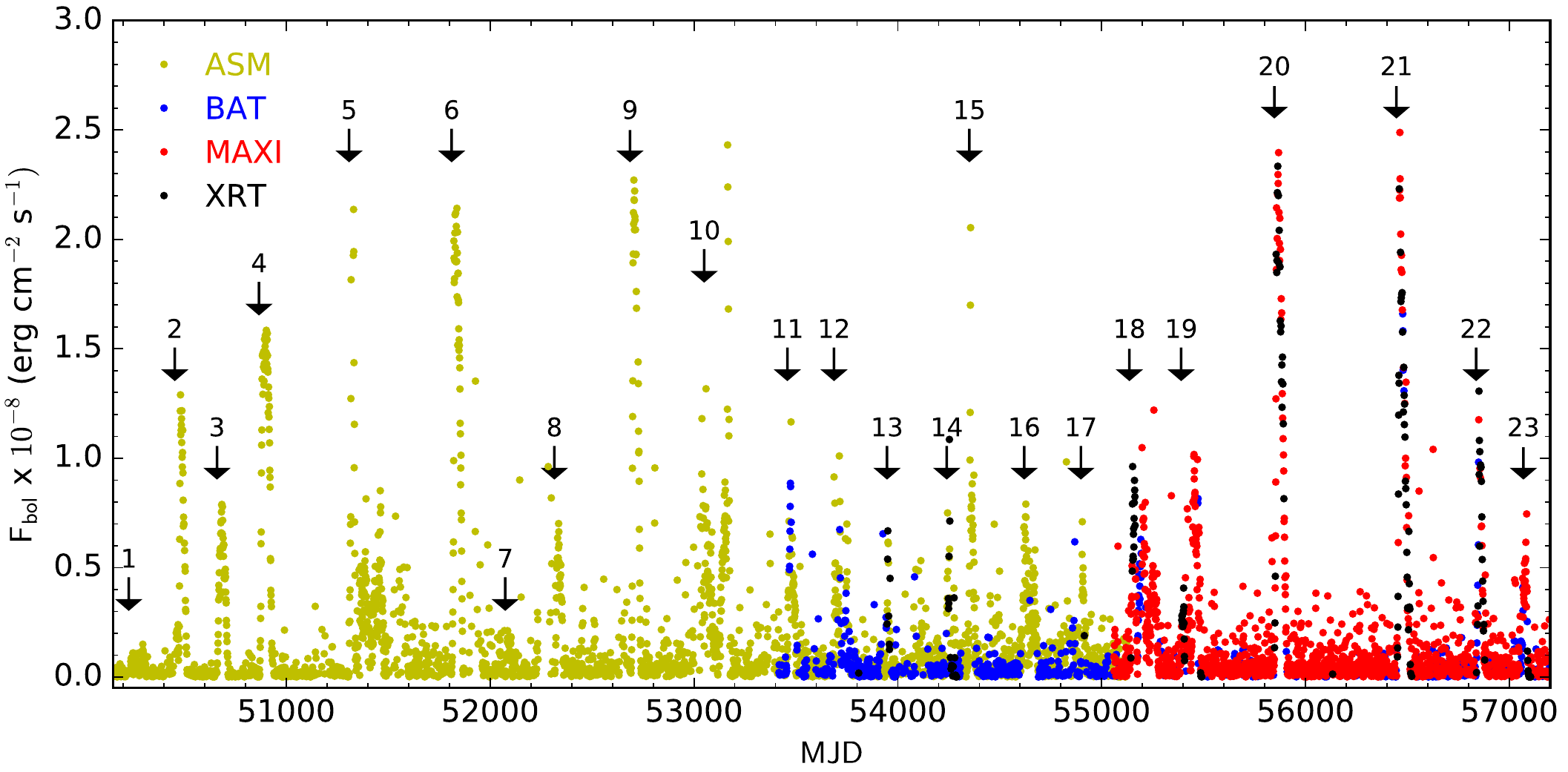}
   \caption{Bolometric flux light curve of Aql X-1 between 1996 and 2015 determined from observations with {\it RXTE}/ASM ($2-10$ keV), {\it Swift}/BAT ($15-50$ keV), MAXI ($2-20$ keV), and {\it Swift}/XRT ($0.5-10$ keV). The arrows indicate the start times of the accretion outbursts detected during this period. We refer to the different outbursts by the numbers allocated in this plot. See Section \ref{fig:Aql_X-1_individual_outbursts_hardsoft} for the light curves of the individual outbursts.}
  \label{fig:light_curve}
\end{figure*}

\subsection{Code input: the 1996-2015 light curve}\label{sec:lightcurve}
\subsubsection{The time-dependent accretion rate} \label{sec:conversion}

Figure \ref{fig:light_curve} shows the bolometric flux of Aql X-1 that we calculated over the period between April 1996 and July 2015. During this period 23 outbursts were detected and monitored with different instruments. From here on, we refer to specific outbursts by their number as indicated in Figure \ref{fig:light_curve}. From the start of the mission of the {\it RXTE} in 1996, Aql X-1 was monitored almost daily with its All Sky Monitor (ASM), with the exception of Sun-constrained windows. By the time of decommission of the {\it RXTE} in 2012, the {\it Swift} mission had been launched (2004) and the MAXI (2009) on board of the {\it International Space Station (ISS)} was activated, providing further coverage of the activity of the source. Besides the {\it Swift} Burst Alert Telescope (BAT) that carries out monitoring observations, the {\it Swift} telescope is equipped with the X-Ray Telescope (XRT), which has been used for many pointing observations of Aql X-1 during different outbursts. Although outbursts of Aql X-1 were observed before 1996 we do not model these, because there were no regular observations during these outbursts. Additionally, there were no continuous monitoring observations between these accretion periods and the start of the ASM monitoring and we therefore do not know if any outbursts have been missed in that period. 

To construct the light curve of Aql X-1 we combined data from the {\it RXTE}/ASM \citep[$2-10$ keV,][]{levine1996}\footnote{\url{https://heasarc.gsfc.nasa.gov/docs/xte/ASM/sources.html}}, the BAT \citep[$15-50$ keV,][]{krimm2013}\footnote{\url{https://swift.gsfc.nasa.gov/results/transients/AqlX-1/}} and XRT \citep[$0.5-10$ keV,][]{evans2007,evans2009}\footnote{XRT light curves of Aql X-1 were obtained from the {\it Swift}/XRT online tool: \url{http://www.swift.ac.uk/user_objects/}} on board {\it Swift}, and MAXI \citep[$2-20$ keV,][]{matsuoka2009}\footnote{\url{http://maxi.riken.jp/star_data/J1911+005/J1911+005.html}}. For the ASM and XRT observations, we averaged multiple pointings taken on one day such that we obtained daily averaged count rate light curves of Aql X-1 for each instrument. In order to combine the data and determine the bolometric flux, one needs to correct for the fact that each of the different instruments operates in different energy ranges and has a different response. Aql X-1 is not constantly bright in each energy range because it has a variable spectral shape (i.e., when the source is in the hard state compared when it is in the soft state). Therefore, we first tried to identify when the source was in the hard and the soft state during each outburst and then determined conversion factors to bolometric flux for each of those. Figure \ref{fig:Aql_X-1_individual_outbursts_hardsoft} shows the individual light curves of the outbursts of Aql X-1 with the determined spectral state indicated.

For outbursts $18-23$ we separate when Aql X-1 was in the hard and the soft state using the hardness ratio determined from MAXI observations. We calculate the hardness ratio as the count rate in the $10-20$ keV band (hard state) divided by the count rate in the $2-10$ keV band (soft state). We define observations for which the hardness ratio is smaller than $0.1$ as  soft state observations and if the ratio is larger than 0.1 as hard state observations. With this method we were able to determine the time of source state transition in outbursts  $18-22$. During these five outburst the source starts in the hard state, but during the rise of the outbursts it transitions to the soft state. For the main extent of the outburst, including the peak, the source stays in the soft state, and only returns back to the hard during the outburst decay. During the full extent of outburst 23, the source was in the hard state. 

For outbursts $1-10$ we only obtained ASM monitoring data and for outbursts $11-17$ from ASM and BAT. Therefore, we cannot use the distinction method based on MAXI data. However, the ASM operated in different energy bands as well and although the bands span a narrower energy range, we can use ASM ratios to make a rough estimate of the spectral state of the source during an outburst. We calculated ASM ratios by dividing the A band intensities ($1.5-3$ keV) by the C band intensities ($5-12$ keV) and define a rough ratio limit of $1.0$. If the ASM ratio during an outburst is primarily larger than $1.0$ we define the outburst as a soft state outburst, and if the ratio is generally smaller than $1.0$ we consider the source to be in the hard state during the outburst. The ASM data are not accurate enough to discriminate state changes during outburst, and therefore we only estimate the spectral state over the entire outburst from ASM ratios. We find that outbursts $1-4$ and $6-9$ are predominantly soft sate outbursts (with possible hard state episodes at the beginning and the end of the outburst, but we are unable to confirm this from the ASM data), while the ratios indicate that during outbursts 5 and 10 the source was always in the hard state. For outbursts $11-17$ we obtained BAT data as well, which enables an additional method of state determination. Since the ASM and BAT instruments observe in respectively soft and hard X-rays, we can separate the different spectral states of the source from the relative intensities. We convert the count rates into Crab intensities using conversion factors of $1 \text{ Crab}=75 \text{ count s}^{-1}$ for the ASM data and $1 \text{ Crab}=0.22 \text{ count s}^{-1} $ for BAT data \citep{krimm2013}. We find that outburst $11-14$, 16, and 17 have similar intensities from ASM and BAT observations, indicating that during these outbursts the source was constantly in the hard state. For outburst 15 we find a drop in the BAT intensities compared to ASM during the middle part of the outburst which we identify as a transition to the soft state. These state identifications are consistent with the ratios determined from ASM observations. 

To convert the count rates measured when the source was in either the hard or the soft state into bolometric fluxes, we use bolometric flux measurements reported in the literature. \citet{king2016} measured the bolometric flux ($0.1-100$ keV) of Aql X-1 in the soft state from simultaneous {\it NuSTAR} and {\it Swift}/XRT observations at MJD  $=56755$ to be $F_{0.1-100\text{ keV}}=1.028\times10^{-8}\text{ erg cm}^{-2}\text{ s}^{-1}$. The XRT intensity at the time of this observation was $231.98\text{ count s}^{-1}$. From this measurement we extract an XRT count rate to bolometric flux conversion factor of $4.4\times 10^{-11}\text{ erg cm}^{-2}\text{ count}^{-1}$. Next, we determined for each day that the source was in the soft state and at which XRT data overlaps with data from one of the other three telescopes a conversion factor from either ASM, BAT or MAXI count rate into XRT count rate. We then calculate for each of the three instruments the average soft state conversion factor for count rate into XRT count rates. We assume that the XRT count rate to bolometric flux conversion factor obtained for MJD $=56755$ holds for all soft state observations and combine this with the count rate conversion factors to obtain for each instrument a soft state count rate to bolometric flux conversion factor (see Table \ref{table:conversionfactors}). 

\begin{table}
\centering
\caption{Calculated count rate to bolometric flux conversion factors in units of $\text{ erg cm}^{-2}\text{ count}^{-1}$ for the hard and soft states of Aql X-1 for each instrument used in this study.}
\label{table:conversionfactors}
\begin{tabular}{lllll}
\hline
                             & {\it RXTE}/ASM     & {\it Swift}/BAT   & MAXI                & {\it Swift}/XRT    \\ \hline
Soft state & $4.5\times 10^{-10}$ & $2.2\times 10^{-6}$ & $8.4\times 10^{-9}$ & $4.4\times 10^{-11}$ \\
Hard state & $1.5\times 10^{-9}$  & $2.7\times 10^{-7}$ & $2.3\times 10^{-8}$ & $1.5\times 10^{-10}$ \\ \hline
\end{tabular}
\end{table}

To calculate hard state conversion factors we use the bolometric luminosities that were obtained from best-fit spectral modelling of {\it Suzaku} observations reported by \citet{zhang2016}. This research reports the analysis of five hard state observation of Aql X-1 for which we have data in our sample taken on the same day (MJD$= 54377, 54382, 54388, 54392, 55852$).  For each date we first convert the reported luminosities back to fluxes (assuming a distance of 5 kpc) and then calculate the count rate to flux conversion factor for the individual instruments. We note that XRT and MAXI observations only coincide with one {\it Suzaku} observation, while ASM and BAT coincide with four and five respectively. We use for each instrument the mean hard state conversion factor calculated (see Table \ref{table:conversionfactors}) for all hard state observations.

Finally, a daily averaged accretion rate is calculated from the bolometric flux \citep[see ][for details]{ootes2016}. We assume the fraction of accreted mass going into X-ray luminosity to be 0.2 and use a distance of $5.0$ kpc. For days with observations from multiple instruments we use data from one instrument in the following order of priority: XRT, MAXI, ASM, BAT. Since the sensitivity of the ASM deteriorated near the end of the {\it RXTE} mission  \citep[see e.g. the performance analysis by][]{vrtilek2013,grinberg2013}, we prioritise BAT over ASM observations for MJD $>55170$.

\subsubsection{Quiescent data}\label{qdata}
We model the temperature evolution of Aql X-1 to fit the cooling data obtained after outbursts 14, 19, 20, 21, and 23. For outbursts 20, 21, and 23 we use the temperatures obtained from {\it Swift}/XRT observations by \citet{waterhouse2016}. After outburst 23, three additional \textit{Swift} observations were taken that were not previously reported. We analysed those observations in the same way as reported by \citet{waterhouse2016}; we refer to that paper for full details. After outbursts 14 and 19 respectively two and one {\it Swift}/XRT observations were available and those were included in our paper in the same way. 

\begin{figure}
	\includegraphics[width=0.99\columnwidth]{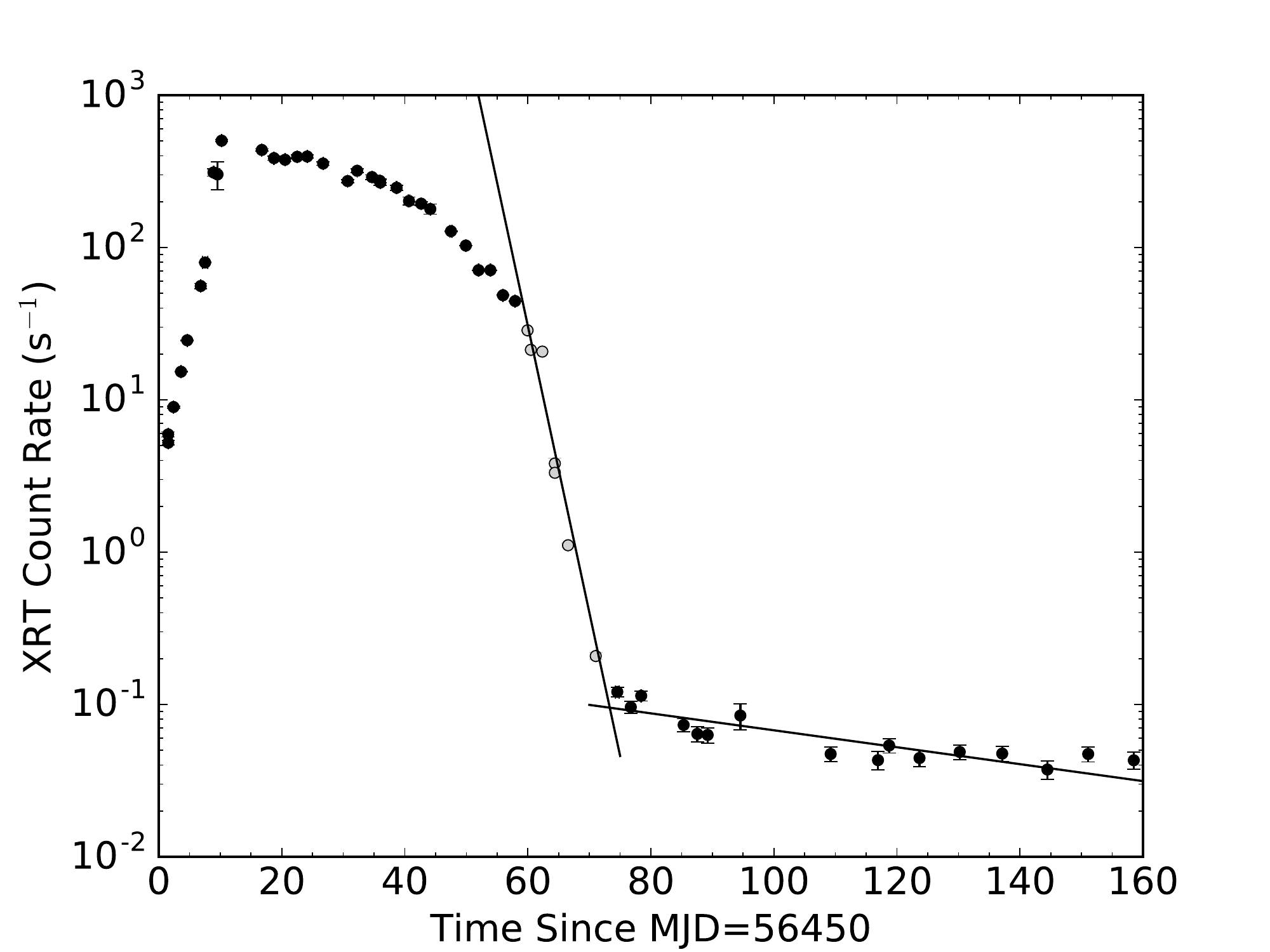}
    \caption{XRT count rate as function of time for outburst 21. The end time $t_0$ of the outburst was determined as the intersection of fits through respectively the outburst decay and the quiescent observations. The grey points indicate the observations that we assume to be part of the outburst decay.}
    \label{fig:decay_fit}
\end{figure}

\subsubsection{Outburst start and end times}
We determine the start times ($t_\text{start}$) of all outbursts as the date that the source becomes detectable according to our obtained light curves (Figure \ref{fig:light_curve}). Similarly, the end dates ($t_0$) of the 18 outbursts for which we have no quiescence {\it Swift}/XRT observations are determined to be the date when the source became undetectable according to our light curves obtained by the ASM, MAXI, and/or BAT. We list these times in Table \ref{tab:outburstproperties}, as well as the accretion rate averaged over the full outburst, $\langle\dot{M}\rangle$, as determined from the daily averaged bolometric flux. The determined start and end times compare well with those reported by \citet{campana2013}

The determination of the end dates of outbursts for which we have {\it Swift}/XRT quiescence observations was done in more detail (i.e. the five outbursts for which we compare the calculated cooling curves with quiescence observations), since the exact end date is of influence for the cooling calculations. For outburst 19 we use the end date calculated by \citet{campana2014}, and for outburst 20, for which the outburst decay was missed, we use the estimated end time reported by \citet{waterhouse2016}. To determine the end date of outbursts 14, 21 and 23, we fitted the decay curve observed with {\it Swift}/XRT as well as the quiescent curve with an exponential decay function and calculated the end date of the outburst as the intersection of the two (see Figure \ref{fig:decay_fit} for the results for outburst 21). We list the $t_0$ values that we determined for those outbursts in Table \ref{tab:outburstproperties} as well. It should be noted that \citet{waterhouse2016} determined $t_0=56518$ as end date for outburst 21 (5 days earlier) and $t_0=57100$ for outburst 23 (4 days earlier) with a similar method. As this is a significant difference for outburst 21 considering the dates of the first quiescence observations (MJD = 56524; either 1 or 6 days after $t_0$, depending on which value for $t_0$ is used), we will compare results that can be obtained with the two different end dates in Section \ref{res:2012outburst}. Elsewhere we will use for outburst 21 the end date that we determined from Figure \ref{fig:decay_fit}. For outburst 23 we do not compare the results for different outburst end times, as for this outburst the first quiescence observation is later in time. This makes the effect of a different end time significantly smaller considering the logarithmic time scale over which we observe crust cooling.

For outburst 20 we also take into account the two reflares shortly after the end of the outburst, first reported by \citet{cotizelati2014}. These started 82 and 208 days after the end of the outburst and last respectively for 32 and 7 days. 

\subsubsection{Envelope composition: upper limits}

\begin{table}
\begin{center}
\caption{Determined outburst properties: start, $t_\text{start}$, and end time $t_0$ of the outburst, time-averaged accretion rate, $\langle\dot{M}\rangle$, as percentage of the Eddington mass accretion rate (with $\dot{M}_\text{Edd}=1.73\times10^{18}\text{ g s}^{-1}=2.74\times10^{-8}M_\odot\text{ yr}^{-1}$), time of the last detected type-I X-ray burst before the end of the outburst, $t_\text{XRB}$, and the calculated upper limit on the helium column depth, $Y_\text{He,max}$, based on the last observed X-ray burst. The outbursts are numbered as in Figure \ref{fig:light_curve}. We also report the year in which the outburst took place.}
\label{tab:outburstproperties}
\begin{tabular}{lllllll}
\hline
\# & year &$t_\text{start}$ & $t_\text{0}$  & $\langle\dot{M}\rangle$ & $t_\text{XRB}^\dag$  & $Y_\text{He,max}$ \\ 
&&&&$\%\dot{M}_\text{Edd}$&&g cm$^{-2}$\\\hline
1 & 1996 & 50232            & 50313          &  \:\;$0.54$             & 50288.4  & $1.4\times10^{9}$             \\ 
2  & 1997 & 50460            & 50514          & \:\;$5.34$              & 50509.0 & $5.5\times10^{8}$             \\ 
3  & 1997 & 50660            & 50716          & \:\;$3.96$              & 50701.5 & $3.0\times10^{9}$             \\
4  & 1998  & 50868            & 50940          & \:\;$8.67$              & 50886.5 & $4.9\times10^{10}$             \\
5  & 1999 & 51308            & 51493          & \:\;$5.15$              & 51439.8 & $1.5\times10^{10}$             \\
6  & 2000 & 51809            & 51873          & $10.32$              & 51856.2 & $5.2\times10^{9}$             \\
7  & 2001 & 52074            & 52123          & \:\;$0.85$              & 52100.8 & $1.6\times10^{9}$             \\
8  & 2002 & 52315            & 52359          & \:\;$3.43$              & 52354.0 & $6.8\times10^{8}$             \\
9  & 2003 & 52686            & 52743          & $10.45$              & 52735.4 & $9.8\times10^{8}$             \\
10 & 2004 & 53029           & 53177         & \:\;$4.86$              & 53153.8 & $2.3\times10^{10}$             \\
11 & 2005 & 53456           & 53507          & \:\;$3.90$              & 53490.9 & $3.0\times10^{9}$             \\
12 & 2005 & 53679           & 53760          & \:\;$3.09$              & 53720.2 & $1.1\times10^{10}$             \\
13 & 2006 & 53941           & 53973          & \:\;$2.06$              & 53962.5 & $7.7\times10^{8}$             \\
14 & 2007 & 54236           & 54288          & \:\;$2.13$              & 54259.2 & $2.6\times10^{9}$             \\
15 & 2007 & 54345           & 54391          & \:\;$4.65$              & 54384.0 & $8.6\times10^{8}$             \\
16 & 2008 & 54603           & 54689          & \:\;$2.91$              & -            & -                    \\
17 & 2009 & 54895           & 54926          & \:\;$2.25$              & -           & -                    \\
18 & 2009 & 55139           & 55282          & \:\;$3.64$              & 55256.2 & $6.8\times10^{9}$             \\
19 & 2010 & 55388           & $55491^*$          & \:\;$3.50$              & 55472.8 & $5.2\times10^{9}$             \\
20 & 2011 & 55843           & 55919         & \:\;$9.87$              & 55905.3 & $9.0\times10^{8}$             \\
21 & 2013 & 56444           & 56523          & \:\;$8.54$              & $56493^\S$             & $8.6\times10^{9}$                   \\
22 & 2014 & 56838           & 56886          & \:\;$4.71$              & $56856^\ddag$ & $1.3\times10^{10}$                     \\
23 & 2015 & 57039           & 57104          & \:\;$1.96$              & -             & -                    \\ \hline

\end{tabular}
\end{center}
$^*$Calculated by \citet{campana2014}\\
$^\dag$ collected from {\it MINBAR}, except for $^\S$ which was reported by \citet{serino2016} and $^\ddag$ reported by \citet{king2016}
\end{table}

It is generally assumed that during a type-I X-ray burst all light elements in the envelope are fused into heavier elements \citep[although see the recent work by][]{keek2017}. The triple-$\alpha$ reaction that ignites the burst fuses the fuel into carbon, after which further nucleosynthesis reactions, depending on the temperature, density, and amount of H present, lead to more proton rich burning ashes \citep[e.g.][]{schatz2001}. This means that over the course of such a burst, the envelope composition changes drastically. After the burst, a new layer of light elements builds as the source continues to accrete. Consequently, for sources that undergo frequent X-ray burst such as Aql X-1, the composition of the envelope at the end of the outburst will depend on how long ago the last X-ray burst occurred. This gives us an opportunity to place upper limits on the amount of light elements in the envelope. 

For each outburst, we search the Multi-INstrument Burst ARchive (MINBAR)\footnote{The MINBAR database, maintained by Duncan Galloway, can be found at \url{http://burst.sci.monash.edu/minbar}.} as well as the literature for the time of the last detected type-I X-ray burst ($t_\text{XRB}$; see Table \ref{tab:outburstproperties}). Based on the average accretion rate in the time between the last X-ray burst and the end of the outburst we calculate the amount of material that is accreted in this period. Assuming a stellar radius $R=11$ km an upper limit on the helium column depth $Y_\text{He,max}$ is derived (see Table \ref{tab:outburstproperties}). It should be noted that this only places very rough upper limits on the envelope composition, since X-ray bursts are easily missed by monitoring telescopes because of their short duration (a few minutes at most). 

Simple models show that for local mass accretion rates $0.1\lesssim\dot{m}/\dot{m}_\text{Edd}\lesssim1.0$, helium will ignite unstably as soon as the accreted layer reaches the thermal instability limit which has been calculated to be at $y_\text{ign}\simeq3\times10^8 \text{ g cm}^{-2}$ \citep{bildsten1998}. Unfortunately, all upper limits are higher than the column depth required to trigger an X-ray burst. We could therefore not use these results as constraints in our models. 

\section{Results}

\subsection{Comparing models}
We fitted the cooling curves of Aql X-1 for a variety of models, taking into account different assumptions for the input parameters. Here we describe each of these models and their results. In all models, the assumed mass is $M=1.6 \text{ M}_\odot$, and the radius $R=11$ km, consistent with the spectral fits of the quiescent cooling data \citep{waterhouse2016}. All 23 outbursts are taken into account in each model, but we allow the parameters for envelope composition and shallow heating to change between the outbursts (unless stated otherwise). For each model the total $\chi^2$ is reduced in order to obtain the best-fit parameters and we calculate $1\sigma$ errors on each of the parameters. It should be noted that for outburst 19 we have only one quiescent observation, and hence the parameters of this outburst cannot be constrained with high accuracy. 

For the 18 outbursts for which we have no cooling observations, we fix the parameters to default values unless described differently. The value for a fixed envelope composition is $y_\text{L}=10^8\text{ g cm}^{-2}$. We assume this is a valid estimate, considering the accreted light element layer required to trigger a type-I X-ray burst ($y_\text{ign}\simeq3\times 10^8\text{ g cm}^{-2}$). The default amount of shallow heating is $Q_\text{sh}=1.5$ MeV nuc$^{-1}$, at $\rho_\text{sh,min}=4\times 10^8\text{ g cm}^{-3}$, which equals the typical amount of shallow heating found for other outbursts. 

\subsubsection{Primary model: a low impurity crust and variable shallow heating and envelope composition}

\begin{figure}
	\includegraphics[width=\columnwidth]{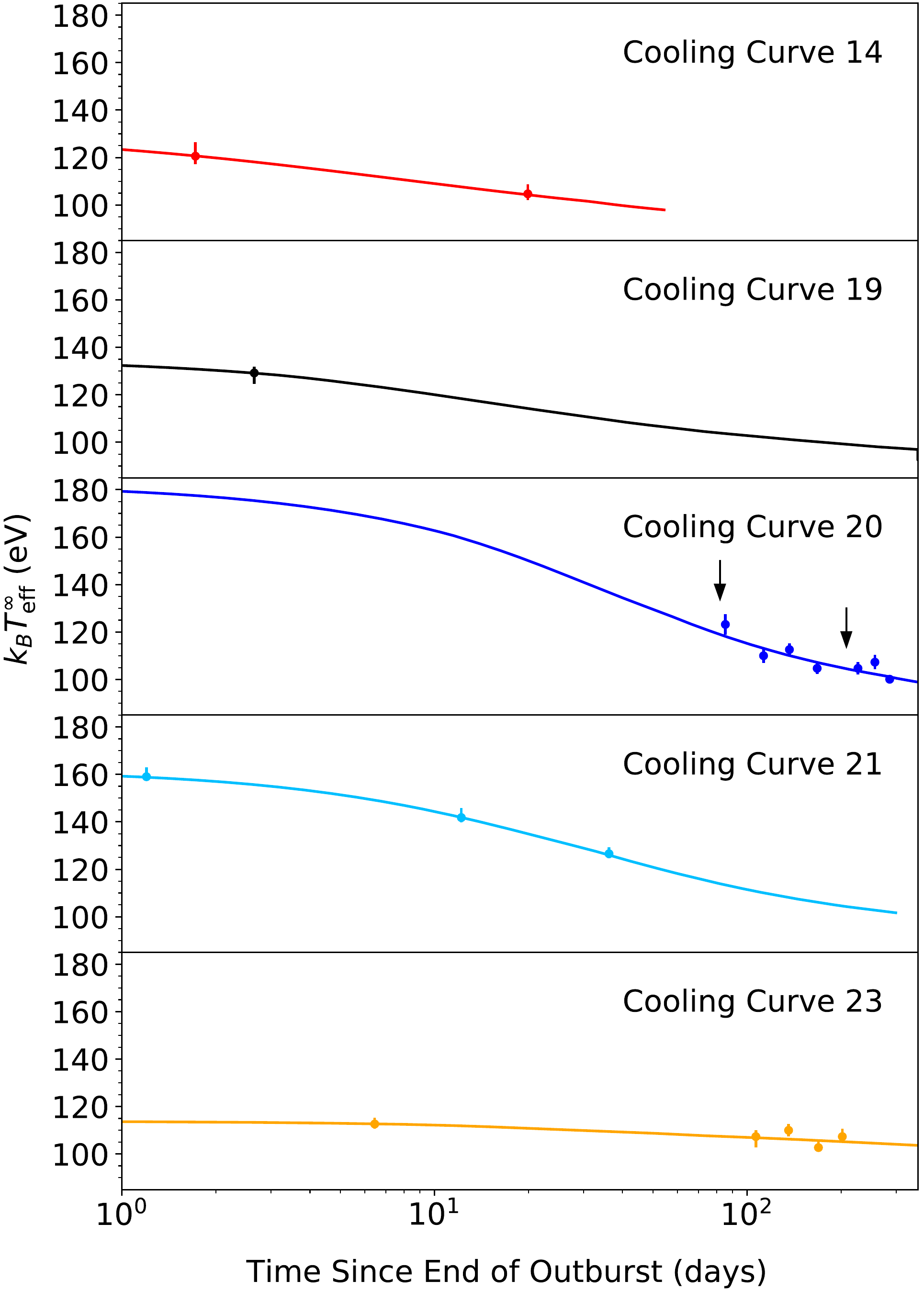}
    \caption{Calculated cooling curves of outbursts 14, 19, 20, 21, and 23 for our primary model of Aql X-1. The amount and depth of shallow heating and envelope composition were changed between different outbursts to reduce the total $\chi^2$ value, while all other input parameters are constant for all 23 modelled outbursts. The arrows in the figure of cooling curve 20 indicate the times of the observed accretion flares \citep{cotizelati2014}. Cooling curves of the remaining 18 modelled outbursts are not shown, because no quiescent observations are available to compare the calculated model with.}
    \label{fig:bestfit}
\end{figure}

\begin{table}
\setlength{\extrarowheight}{0.1cm}
\begin{center}
\caption{Fit parameters with $1\sigma$ errors for the primary model. In this model the crust impurity parameters are fixed to 1.0. We fitted for the initial core temperature $\tilde T_0$ and the shallow heating and envelope parameters. For the five outbursts with quiescent observations the shallow heating and envelope parameters were kept untied between outbursts in the fit. For the other 18 outbursts these parameters were fixed to canonical values (indicated by the values without errors). } 
\label{tab:bestfit}
\begin{tabular}{lllll}
\hline
Outburst & $\tilde T_{0}$& $Q_\text{sh}$  & $\rho_\text{sh,min}$ & $\log(y_\text{L})$        \\
           &  $\times10^7$ K&  \mev             & $\times10^9$ g cm$^{-3}$            &  g cm$^{-2}$                        \\ \hline
14       & & $2.9^{+1.6}_{-1.2}$           & $0.1^{+1.3}_{*}$        & $8.2^{+1.2}_{*}$       \\
19       & & $1.3^{+1.4}_{-1.0}$           & $0.3^{+35.9}_{*}$        & $8.8^{*}_{*}$          \\
20       & & $3.7^{+1.5}_{-0.9}$           & $0.4^{+7.9}_{*}$        & $8.3^{+0.7}_{-0.9}$ \\
21       & &$2.3^{+0.5}_{-0.3}$           & $0.4^{+0.7}_{*}$        & $8.8^{+1.1}_{-1.5}$  \\
23       & &$0.9^{+0.8}_{-0.6}$           & $2.8^{+13.7}_{*}$        & $9.8^{*}_{-2.5}$     \\
Other  & $8.9^{+2.3}_{-1.5}$ &$1.5$           & $0.4$      & $8.0$                     \\ \vspace{-0.35cm} \\
\hline
\end{tabular}
\end{center}
$^* $ The error exceeds the maximum or minimum allowed value of the parameter in our model.
\end{table}

\begin{figure*}
	\includegraphics[width=\textwidth]{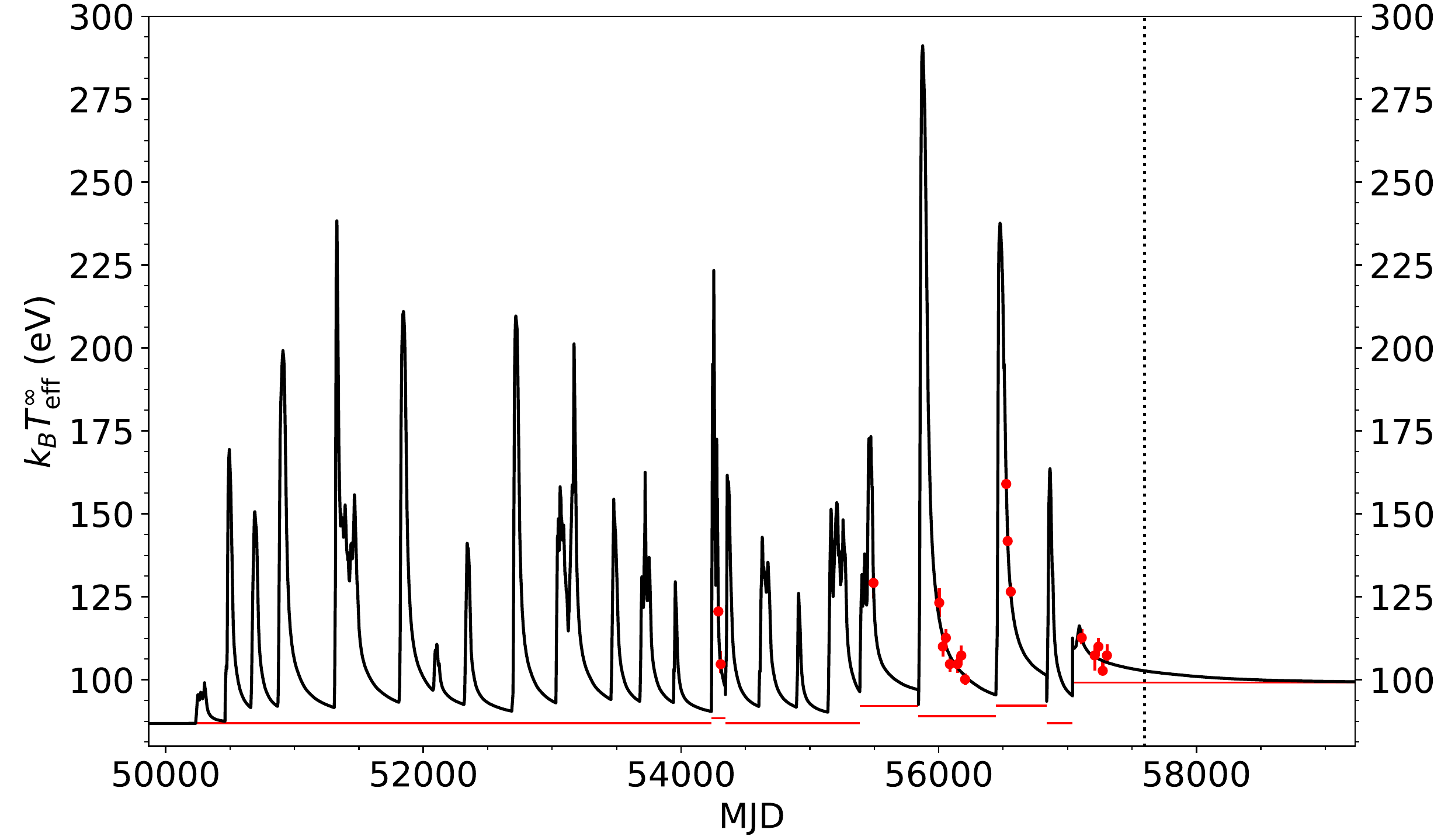}
    \caption{Calculated effective temperature as function of time during all 23 modelled outburst, for the primary model assuming a low-impurity crust. The red points indicate the quiescent observations during the cooling curves of outburst 14, 19, 20, 21 and 23 (see Figure \ref{fig:bestfit} for a close-up of those cooling curves). The red lines represent the observable base level of the source for the envelope composition assumed at that time; i.e. the effective temperature to which the source would cool down if the recurrence time was long enough to restore crust-core equilibrium. The discontinuities in both the calculated effective temperature and the base levels are caused by the fact that we allow an instantaneous change in envelope composition at the start of a new outburst. The envelope composition remains constant during the course of an outburst and subsequent quiescent period. The vertical dotted line indicates the start of the 2016 outburst \citep{sanna2016a,sanna2016b}, and shows that the source needs more time to cool down to the base level than the recurrence time.}
    \label{fig:full_bestfit}
\end{figure*}
Figure \ref{fig:bestfit} shows the calculated cooling curves for the model in which we leave the core temperature free and allow the envelope composition and amount and depth of shallow heating to vary per each of the five outbursts with cooling data. The best-fit parameters are shown in Table \ref{tab:bestfit} and also as model 1 in Table \ref{allmodelparams}. In this model we kept the impurity parameter of all density regions in the crust fixed at $Q_\text{imp}=1$. We investigated various possibilities for the impurity parameter of the crust by changing the input parameters by hand, but found that none of the cooling curves showed indications for an inner or outer crust that is either more or less conductive than the canonical value of $Q_\text{imp}=1$. We will refer to the model described in this section as the primary model.

Matching the calculated cooling curves with the observations requires to find the right combination of the amount and depth of shallow heating and envelope composition for each individual outburst, while taking into account that the core temperature remains constant for all outbursts. We find that for outbursts 14, 20, and 21 we need a significant amount of shallow heating  of $2.3-3.7$ \mev at shallow depths ($1-4 \times 10^8 \text{ g cm}^{-3}$) to explain the steep cooling curves (see Figure \ref{fig:bestfit}). Cooling curve 23 has a significantly shallower temperature evolution and consequently only $0.9$ \mev is needed at a larger depth of $2.8\times 10^9\text{ g cm}^{-3}$ to explain these observations. In this best fit model, we find $Q_\text{sh}=1.3$ \mev for outburst 19.

Assuming that the neutron star has a high conductivity crust (low impurity), it needs $\sim1500$ days to cool back to crust-core equilibrium as can be seen from the last outburst in Figure \ref{fig:full_bestfit}. The observed base level corresponding to crust-core equilibrium depends on the core temperature and the envelope composition. Since the cooling observations only reach to $\sim300$ days into quiescence and the source does not show quiescent episodes lasting multiple years in the observed epoch, we do not know what the base level of Aql X-1 is. From our model we find that the data is best fitted for a core temperature of $\tilde T_0=8.9\times10^7$ K, in combination with light element column depths reaching from $y_\text{L}=1.6\times10^8$ g cm$^{-2}$ to $y_\text{L}=6.3\times10^9$ g cm$^{-2}$ for the different outbursts with cooling data corresponding to observed base levels of $\sim 88-99$ eV. Since the envelope composition determines the conversion from boundary temperature to effective temperature, varying this parameter shifts the complete calculated cooling curve to higher or lower temperatures and thus also the base level (see Figure \ref{fig:Tb-Ts}). This means that even though the core temperature does not vary, the base level does when the envelope composition is assumed to change between outbursts \citep{brown2002}. Figure \ref{fig:full_bestfit} shows the results of the full model: the calculated effective temperature as function of time during all outbursts and subsequent cooling phases. The red line indicates the base level to which the source will cool down if it restores crust-core equilibrium, which is determined by the core temperature in combination with a variable envelope composition. Because we vary the envelope composition per outburst, the base level goes up or down between outbursts. From this figure one can clearly see that the source does not have enough time between outbursts to cool down to the base level.

The two reflares at 82 and 208 days after the end of outburst 20 were taken into account in the primary model. We assumed for the reflares the same conditions as during the main outburst, but kept the shallow heating strength fixed at 0 MeV nuc$^{-1}$. The reflares show no effect on the cooling curve (Figure \ref{fig:bestfit}). If shallow heating is turned on during the reflares at the same strength as during the main outburst, no change in $\chi^2$ is observed. 

\subsubsection{Constant envelope composition}
To match the calculated cooling curves with the observations we varied the envelope composition for different outbursts in the primary model. We also created a model in which we kept the envelope composition tied between all 23 outbursts. The envelope composition was allowed to change, but had to remain constant between all outbursts. The other free parameters were the core temperature and shallow heating strength and depth (both of which were untied). 

The $\chi^2$ value obtained when keeping the envelope composition constant between outbursts was slightly higher compared to the primary model (see model 2 in Table \ref{allmodelparams}). The differences in steepness and initial surface temperature of the cooling curve between outbursts  can be (partly) compensated for by varying the other fit parameters. However, in order to keep the envelope composition tied between outbursts, a larger difference in shallow heating parameters between the outbursts is required. Specifically, for outburst 23 the required shallow heating increases to 16.4 \mev at unusual high density ($\rho_\text{sh,min}=1.7\times 10^{10}\text{ g cm}^{-3}$).

\subsubsection{Constant shallow heating parameters}
The fit values for the amount and depth of shallow heating for the different outbursts in the primary model are similar to the typical values found in crust-cooling systems that require shallow heat. However, since the origin of the shallow heat has not yet been constrained, it is unclear how much the amount of shallow heating differs either per source, per outburst, or even during an outburst. \citet{deibel2015} and \citet{parikh2018} showed that the outbursts of MAXI J0556-332 are best-fitted with different shallow heating parameters. One possibility to explain variations in shallow heat from different outbursts of one source is that the shallow heating might be related to the accretion flow configuration as proposed by \citet{zand2012}. We tried therefore a variety of models to investigate different possibilities for shallow heating and compared the obtained calculated cooling curves with those obtained using the primary model.

First we created a model in which we kept all shallow heating parameters (both strength and depth) tied between all outbursts. The obtained model has a significantly higher $\chi^2$ than the primary model (see model 3 in Table \ref{allmodelparams}). This is due to the fact that cooling curves 14, 20, and 21 have significantly steeper slopes and a larger observed temperature drop than cooling curve 23, which is difficult to account for when using a constant amount of shallow heating. Changing the envelope composition between outbursts cannot compensate for this effect, because this shifts the entire cooling curve while leaving the steepness unaffected. We then also tried two models in which we tied respectively only the shallow heating strength (model 4) and depth (model 5) between the outbursts with cooling observations. Model 4 has a slightly higher $\chi^2$ and model 5 a similar $\chi^2$ compared to the primary model. 

Second, we created a model in which we could turn off shallow heating when Aql X-1 is either in the hard or the soft state during an accretion outburst. During outbursts 14 and 23 Aql X-1 was observed to be in the hard state throughout the entire outburst, while during outbursts 19, 20, and 21 Aql X-1 was mainly in the soft state, except for a small part of the rise and decay of the outburst (see Figure \ref{fig:Aql_X-1_individual_outbursts_hardsoft}). First we created models in which we kept the amount and depth of shallow heating (activated only when the source was in either the hard or soft state) tied between outburst. However, neither a good fit of the data can be obtained in this way for the hard state shallow heating assumption nor for the soft state shallow heating assumption. Therefore we then tried models in which the amount of hard/soft state shallow heating is allowed to be different per outburst.

Turning on shallow heating only when the source is in the soft state allows us to obtain good agreement between the calculated cooling curves and observations for outbursts 19, 20, 21, and 23, as long as the amount of shallow heating is allowed to vary per outburst. However, this model cannot reproduce the data of cooling curve 14, because during this outburst the source only resides in the hard state. Obtaining a $20$ eV drop in the first 20 days after the end of the outburst is not possible without shallow heating. This problem does not arise for outburst 23, because that cooling curve is so shallow that it can be modelled without shallow heating. 

If shallow heating is turned on only when the source in the hard state, a model can be obtained that provides a satisfactory fit of the observations for all outbursts. However, outbursts 20 and 21 have rather significant temperature drops in their cooling curves while the source is in the hard state for a very short amount of time during these outbursts. On the other hand, outburst 23 has a very shallow cooling curve while the source is in the hard state for a long time. Consequently, to obtain good quality fits for all cooling curves, the difference in the amount of shallow heating between outbursts has to be a factor $\sim10$ when only allowing for shallow heat when the source is observed to be in the hard state.

\subsubsection{Low conductivity pasta layer}\label{res:pasta}
In the primary model we did not take into account the effect that a pasta layer deep inside the inner crust of a neutron star can have on the temperature evolution during and after outbursts. A pasta layer might have a high impurity factor \citep{pons2013,horowitz2015} that can slow down the heat flow to the core and can hence allow a large difference in temperature between the core and the crust. This will influence the cooling curve in last phase before it reaches the base level \citep[see e.g.][]{deibel2016}. To test how a pasta layer would influence our obtained results, we created a model that includes a high impurity layer (with either $Q_\text{imp}=40$ or  $Q_\text{imp}=100$ \citep{horowitz2015,pons2013}) representing the pasta phase in the crust at a depth $\rho>8\times 10^{13} $ g cm$^{-3}$. Outside the pasta layer, the crust is assumed to be highly conductive ($Q_\text{imp}=1$). 

When a pasta layer is included in our model, very little influence on the calculated cooling curves is observed if we use the same input parameters as for the primary model (Table \ref{tab:bestfit}). All cooling curves lay $\sim 1-2$ eV above those of the primary model, due to the fact that heat can flow less easily into the core. However, the increase in $\chi^2$ value can easily be compensated for by adjusting other parameters (for example, the envelope composition). As the maximum duration of the quiescence period between the modelled outbursts ($\sim 500$ days) is shorter than the thermal time scale of the deepest layers of the crust, the observed cooling curves never reach the phase where they are sensitive to the conductivity of the pasta layer. We can therefore not constrain the presence of a low conductivity pasta layer in Aql X-1. 

\subsection{Outburst 21: determining $t_0$ and modelling the outburst }\label{res:2012outburst}
Since we used a different start time $t_0$ for the quiescent phase of outburst 21 than \citet{waterhouse2016}, we made additional models using their end time to test if we can obtain a similarly good fit of the data if we use their quiescent start time. With this end date, the otherwise last decay point becomes part of the quiescence observations resulting in an extra point that is considered to be part of the cooling period. This observation is only two days after the end of the outburst (that is assumed by \citet{waterhouse2016}) and the temperature measured  is $182$ eV. This is significantly higher than the initial temperatures of any of the other cooling curves and requires a steep decay over the first $\sim 40$ days of quiescence.

\begin{figure}
	\includegraphics[width=\columnwidth]{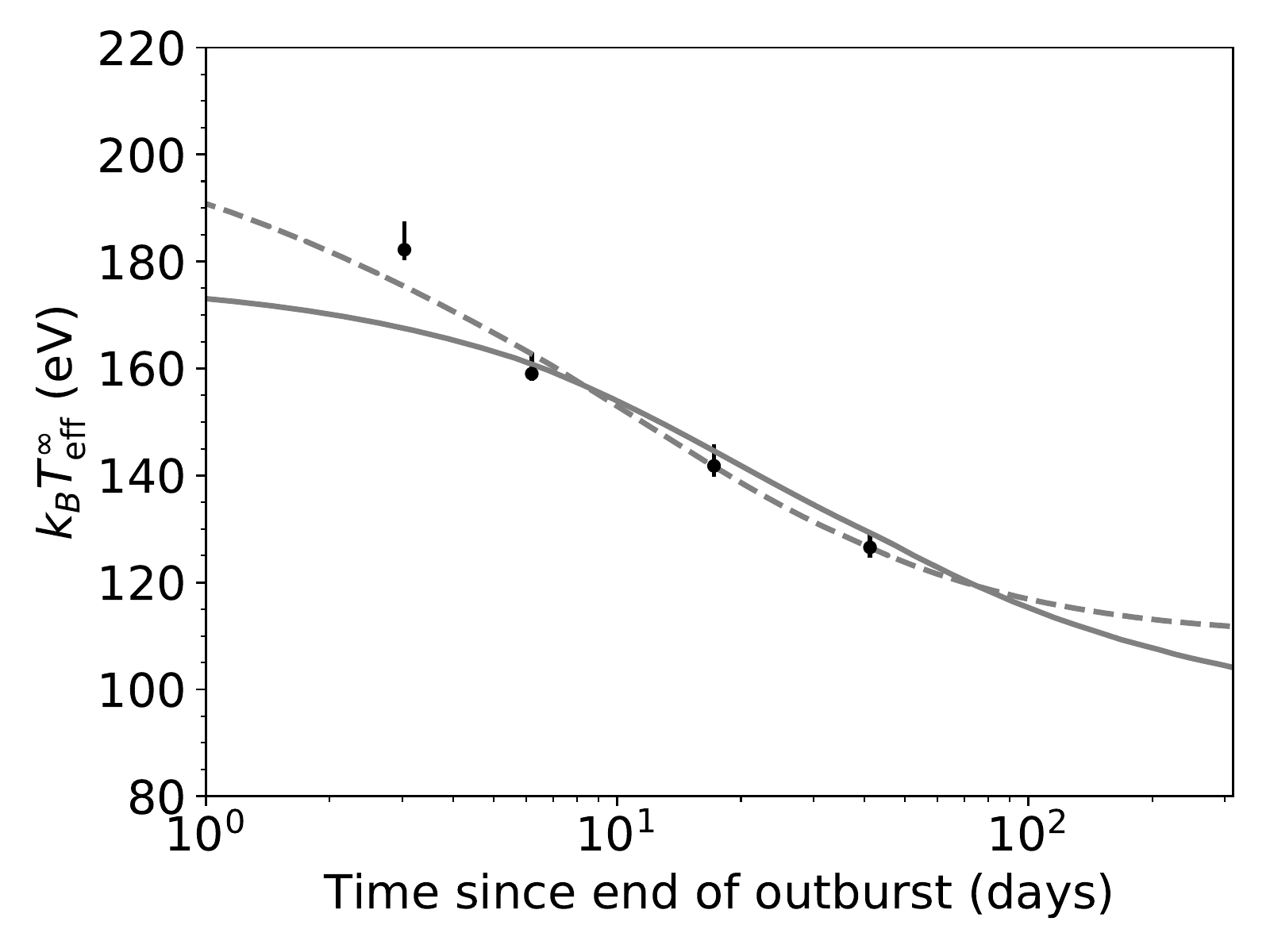}
    \caption{Best-fit calculated cooling curves for outburst 21 when using the end time of the outburst calculated by \citet{waterhouse2016}, instead of the end time that is calculated in this work. The first model assumes constant accretion rate during the outburst and does not take into account previous outbursts (dashed curve) as in \citet{waterhouse2016}, while the second model uses the method described in this work (solid curve). Note that for each of the models we show the best-fit of the data and therefore the models have different input parameters.}
    \label{fig:2012_models}
\end{figure}

Besides the difference in $t_0$, the study performed by \citet{waterhouse2016} differs from ours in the fact that it does not take into account previous outbursts, and it uses a constant mass accretion rate during the outburst. Moreover, the constant accretion rate assumed during outburst 21 ($\dot{M}=0.23 \dot{M}_\text{Edd}$) is higher than the time-averaged accretion rate that we determined ($\langle\dot{M}(t)\rangle=0.10\dot{M}_\text{Edd}$ if we assume their outburst start and end time). Finally, the outburst duration that \citet{waterhouse2016} assumed (0.15 yrs) is shorter than the duration that we determined (0.21 yrs). Therefore, we made two models that both assume the end time for outburst 21 determined by \citet{waterhouse2016}: one that is otherwise the same as the primary model, and one that uses a constant mass accretion rate equal to that of \citet{waterhouse2016} for a duration of 0.15 yrs, without taking into accretion previous outbursts. The second model thus uses the same assumptions as made by \citet{waterhouse2016}. The effects of taking into account previous outbursts will be discussed in Section \ref{sec:historyinfluence}.

Figure \ref{fig:2012_models} shows the best-fit calculated cooling curves for these two models. It should be noted that the best-fit parameters of the two models are different. It is evident that a reasonable fit through all four observations is only possible if a constant accretion rate is assumed during the outburst (dashed curve), although this fit still undershoots the first quiescence observation. In this case the crust remains hot until the end of the outburst and consequently has a relatively steep decay. On the other hand, for the model that uses a time-variable accretion rate based on the observed light curve (solid curve), the temperature profile in the crust changes during the outburst decay; the outer layers of the crust already cool before the end of the outburst due to the decreasing accretion rate. Consequently, the calculated cooling curve is more convex compared to that of the model that assumes a constant accretion rate and we were not able to calculate a cooling curve for this model that is steep enough to reasonably fit all four observations. 

\section{Discussion}

\subsection{Modelling multiple outbursts}
We extended our neutron star crust cooling model {\tt NSCool} \citep{page2013,ootes2016} to model the accretion rate variability during multiple accretion outbursts based on the observed light curves and applied this to Aql X-1, for which 23 outbursts have been observed between 1996 and 2015. After five of those observed outbursts, {\it Swift}/XRT quiescence observations have been obtained that can be interpreted as cooling of the accretion-heated neutron star crust \citep{waterhouse2016}. We thus assume that the observed intensity decay in quiescence is caused by crust cooling and not by accretion phenomena (with the exception of the two observed accretion flares after outburst 20). Aql X-1 serves as a test source to investigate the effects of the accretion history on current outbursts in transients with short outbursts and recurrence times. We fitted the cooling curves of the five outbursts with cooling data to look for the best-fit crustal parameters, while taking into account the full accretion history. In the fitting process we investigated the possibility that some of the crustal parameters change between different outbursts. 

We found that the calculated cooling curves compare well with the observations if we assume a high thermal conductivity crust ($Q_\text{imp}=1$) and if we vary per outburst the envelope composition of the crust and the amount and depth of heat generated by the shallow heating mechanism. A model in which all parameters are tied between the outbursts does not reproduce the data well (see model 6 in Table \ref{allmodelparams}). The amount and depth of the shallow heating are similar to the values found for other crust cooling sources. Most importantly, we find that Aql X-1 does not have enough time between outbursts (250 days on average, with a maximum of $\sim500$ days) to cool down to the crust-core equilibrium state (the base level of the cooling curve, which depends on the initial core temperature and the (variable) envelope composition). For this reason we cannot constrain the base level of this source (the required quiescence period would have to be $>10^3$ days to observe this). The presence of a potentially high-impurity (low conductivity) pasta layer in the inner crust remains unconstrained as well. Effects of such a layer can only be observed a few thousand days after the start of quiescence, which is much longer that the outburst recurrence time.  

Outbursts 20 and 21 were previously fitted by \citep{waterhouse2016}. Comparing outbursts 20 and 21 from our primary model with the results of \citet{waterhouse2016}, we find similar results for the shallow heating strength and also find that the best-fit crustal parameters of the two outbursts are consistent with each other within the $1\sigma$ error. We made an additional model in which we tied all parameters of outburst 20 and 21 (model 7 in Table \ref{allmodelparams}) which confirms that the two remarkably similar outbursts can be modelled with the same assumptions for the microphysics in the crust.  Consistently with \citep{waterhouse2016}, from our cooling calculations we predict the base level of Aql X-1 to be $88-99$ eV. However, different from that research, we used time-dependent accretion rates during the outburst calculated from the observed light curve and take into account the effects of previous outbursts on the thermal state of the neutron star crust. Also, we use a different end time for outbursts 21 and 23 (see Section \ref{sec:discussion-endtime}).

\subsubsection{Constraining crustal parameters from multiple outburst modelling}
Modelling in detail the effects of the full outburst history on the thermal state of the neutron star crust provides advantages towards constraining the microphysics of the crust compared to modelling only one outburst. \citet{waterhouse2016} showed that Aql X-1 has two outbursts that are so similar in both outburst and quiescence behaviour, that their cooling curves can be combined and modelled as one. But even for outbursts where this is not the case, it is advantageous to model all observed outbursts collectively; by which we mean that we model the full accretion and quiescence history (chronologically) in one run of the code. 

Whereas some parameters might change between outbursts (such as composition of the envelope and likely the strength and depth of the shallow heating), others must be constant: mass, radius, and initial core temperature. In this study we have assumed that the lattice impurity parameter (which sets the thermal conductivity) of different layers of the crust is constant between outbursts. Modelling multiple cooling curves collectively while keeping some parameters constant will reduce the degeneracy between input parameters that exists in cooling models. While for one cooling curve very similar results can sometimes be obtained with different combinations of parameters, this degeneracy is partly lifted when modelling multiple cooling curves. This is enhanced by the fact that with more cooling curves of the same source, the sampling of each part of the cooling curve (corresponding to the thermal properties of specific layers of the crust) will increase. Although there is not one constant underlying cooling curve for all outbursts, at similar times in different cooling curves, the same layers are probed. This means that if one cooling curve provides multiple possibilities for some crustal parameter, these possibilities might be further constrained from another cooling curve if it has better observational sampling around the time that the effects of that parameter can be observed.

\begin{figure*}
	\includegraphics[width=0.975\columnwidth]{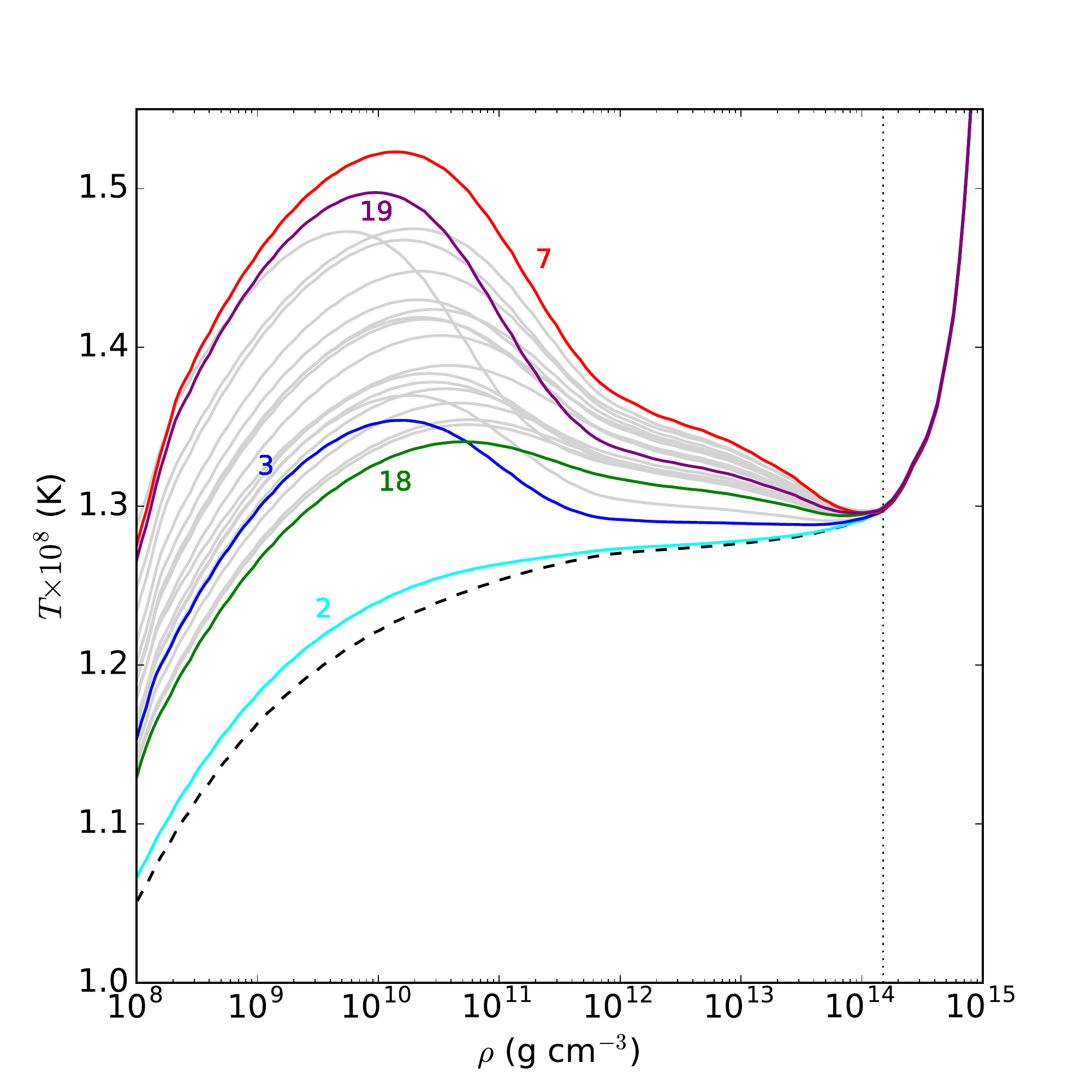}\qquad\qquad\includegraphics[width=0.975\columnwidth]{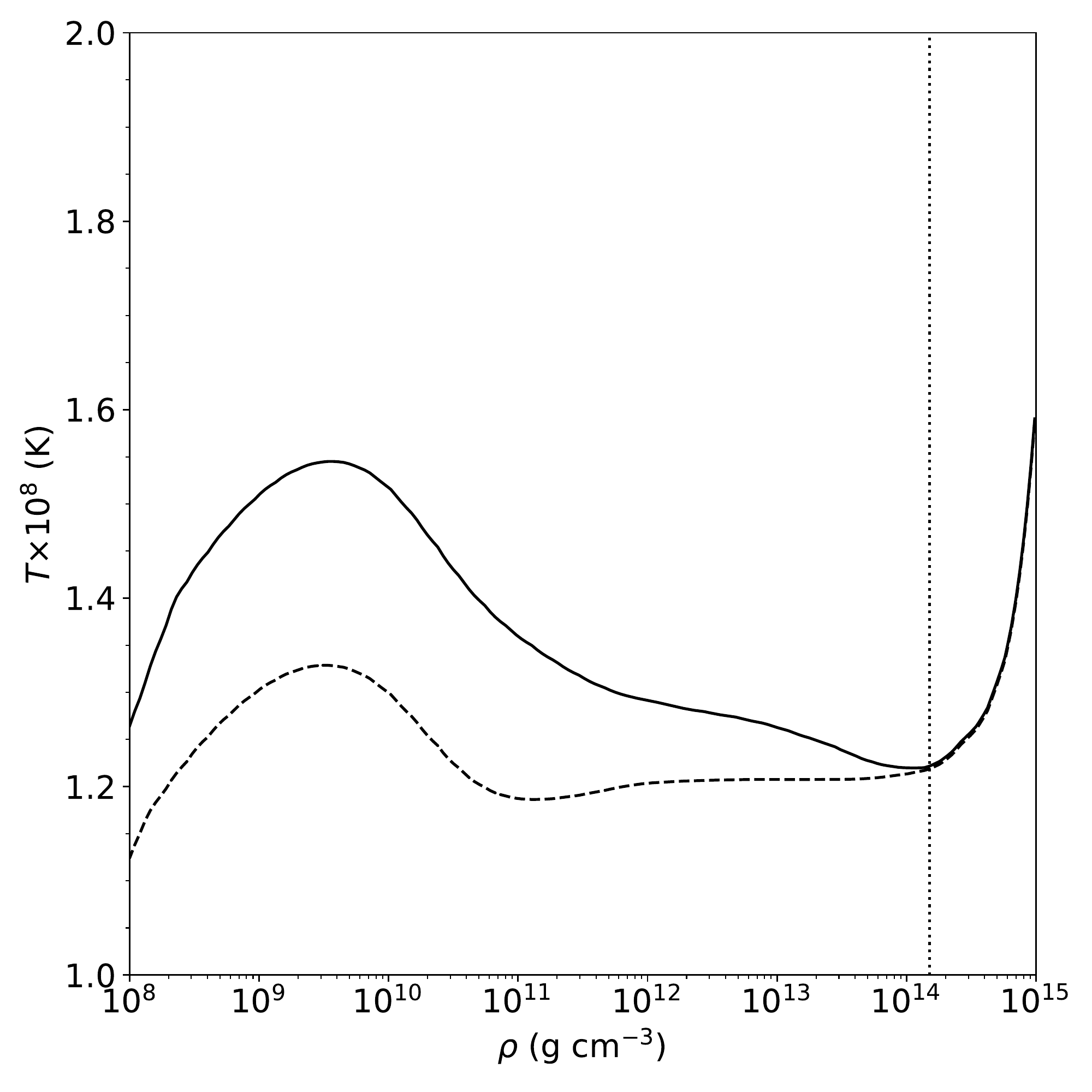}
    \caption{Left: temperature profiles of the crust of the neutron star in Aql X-1 at the start of each outburst as obtained with the primary model. The dashed curve is the temperature profile at the start of outburst 1, when the crust and the core are in thermal equilibrium. The coloured numbers indicate the outburst numbers. Right: temperature profile at the end of outburst 23 with (solid curve) and without (dashed curve) taking into account the full accretion history during the 22 previous outbursts. Both panels use the crustal parameters as determined from the primary model. Note that both panels show the local, non-redshifted temperature (a plot of the red-shifted, $\tilde T$ internal temperature would show a uniform $\tilde T$ in the core). The dashed vertical line indicates the crust-core transition. }
    \label{fig:Tprofiles}
\end{figure*}

In the case of Aql X-1 it was challenging to model the differences in initial temperatures (the temperature observed around the start of the cooling curve), and steepness (cooling rate) between the different cooling curves (see Figure \ref{fig:bestfit}). These differences are partly caused by differences in the outburst properties (e.g. duration, mass accretion rate, and outburst decay rate), but otherwise depend on the input parameters of the model. Outbursts 14 and 23 are both $\sim 120$ eV at the start of the cooling curve, but cooling curve 14 shows a 10 eV drop in temperature in only 10 days, while cooling curve 23 only shows a $\sim5$ eV decay in temperature over more than 100 days. On the other hand, cooling curve 21 starts out very hot at $159$ eV one day after the start of quiescence, and cooling curve 20 shows a decay in temperature at late times ($\gtrsim100$ days into quiescence) when the cooling curve of outburst 23 seems roughly flat.

Outbursts 20 and 21 had larger overall accretion rates than outbursts 14 and 23. Due to its high outburst accretion rate the high initial temperature of outburst 21 (159 eV) could be reproduced with a slightly smaller amount of shallow heating compared to outburst 14, even though after this outburst the source was observed at a lower temperature of 120 eV. Because of the higher accretion rate during outbursts 20 and 21, a lot of heat is stored in the outer crust when using a significant amount of shallow heat (2.3-3.7 \mev in this case). This heat (that comes on top of the residual heat from the previous outbursts) can account for the late time-decay in temperature of the cooling curve corresponding to that outburst. 

To overcome the differences in the cooling trends while using the same initial core temperature, outburst 14 had to have significantly stronger shallow heating than outburst 23 to obtain the observed steeper initial decay, in combination with a lower light element column depth to reduce the initial temperature to a similar level between the two. Similar arguments hold when one compares the cooling curves of outbursts 20 and 23. During these cooling periods the source has a similar observed temperature at $\sim200$ days in to quiescence, while for cooling curve 20 the cooling trend at that time is much stronger than for curve 23 (comparing the observations of the two cooling curves $\sim100-300$ days in quiescence). This indicates that cooling curve 20 might have a lower base level than cooling curve 23 and hence that, apart from stronger shallow heating during the outburst, the two are best modelled with different envelope compositions (i.e. cooling curve 20 must have a lower light element column depth). This is also what we find in the primary model, although the envelope composition of outburst 23 is not well-constrained. All in all, despite the limited coverage of the quiescence episodes of Aql X-1, reasonable constraints can be obtained when modelling all outbursts collectively.

\subsubsection{The influence of the outburst history on the current outburst and cooling calculations}\label{sec:historyinfluence}
By modelling the full outburst history of Aql X-1 collectively we showed that with the assumed crustal properties, the source does not have enough time between outbursts to cool down to its base level (Figure \ref{fig:full_bestfit}). This means that the crust-core equilibrium is not fully restored after an outburst before the start of the next one. The left panel in Figure \ref{fig:Tprofiles} shows the temperature profiles in the neutron star crust at the onset of each outburst for the primary model. The dashed black curve is the temperature profile at the start of the first outburst and hence shows the temperature profile when the crust and core are in equilibrium.  

The first outburst is rather faint and has a low time-averaged accretion rate (see Figure \ref{fig:light_curve}, Table \ref{tab:outburstproperties}), but nevertheless at the onset of the second outburst (147 days after the end of the first outburst) the temperature profile (cyan curve in Figure \ref{fig:Tprofiles}) is elevated compared to crust-core equilibrium. If all outburst and recurrence times were similar, this would have an enhancing effect in the sense that at the start of each consecutive outburst the crust will be a little bit hotter. However, this is not the case for our model of Aql X-1 since the recurrence times, accretion rates, outburst profiles, and the shallow heating differ per outburst. As can be seen from Figure \ref{fig:Tprofiles}, for each new outburst the initial temperature profile is different. At the start of outburst 19 the outer crust is still very hot because outburst 18 had a relatively long duration and the quiescence period between the two outbursts is one of the shortest (106 days). On the other hand, at the onset of outburst 18 the temperature profile is not extreme, because the preceding outburst was short (31 days) and the quiescence period after both outburst 16 and 17 were relatively long (206 and 213 days respectively), allowing much heat to flow out of the crust (into the core and also from the surface). Yet the crust is hottest at the onset of outburst 7, which was preceded by an outburst with average outburst duration (64 days) and quiescence period (201 days), but which had a high time-averaged accretion rate ($\langle\dot{M}\rangle=$0.10 $\dot{M}_\text{Edd}$). 

The right panel of Figure \ref{fig:Tprofiles} visualises the consequences of taking into account previous outbursts in our crust cooling model. Here we show the temperature profile at the end of outburst 23 for the primary model taking into account all preceding outbursts (solid curve) and the temperature profile assuming the same crustal parameters, but without taking into account the accretion history (dashed curve). A significant difference in temperature at all layers of the crust can be observed resulting in a difference in observed surface temperature of $\sim 6$ eV at the onset of the cooling curve. Moreover, the decay trend at later times in the cooling curve is also affected due to the significant difference in the amount of heat stored in the deeper layers of the crust. This shows that for sources with short recurrence times, such as Aql X-1, taking into account the accretion history will have a strong effect on the exact values of the inferred crustal parameters of the neutron star. 
	
\subsection{Variation in parameters between outbursts}

\subsubsection{Variable envelope composition}
The envelope composition was allowed to vary between outbursts in the primary model. When the envelope composition was tied between outbursts the $\chi^2$ increased slightly, and to reproduce the data the shallow heating of outburst 23 had to be unusually high and deep. This suggests that 
 the envelope composition at the end of each outburst is different. 
 
A changing envelope composition between outbursts is supported by our current understanding of the envelopes of accreting neutron stars \citep{brown2002}. The fact that type-I X-ray bursts are observed for this source indicates that there is constant buildup of a layer of light elements that is explosively fused into heavier elements once the layer reaches a column depth at which it becomes thermally unstable. The envelope composition at the end of an outburst would then depend on how much matter has been accreted since the last X-ray burst and likely this amount of matter varies significantly for each outburst. 
 
\subsubsection{Variable shallow heat?}
In our primary model, we found that the strength of shallow heating varied per outburst between 0.9 and 3.7 \mev and the minimum depth between $10^8-3\times 10^9\text{ g cm}^{-3}$. Although the 1$\sigma$ errors on the shallow heating depth in this model do overlap, this is not the case for the shallow heating strength. A model in which we kept both shallow heating parameters tied between outbursts confirms that the data cannot be reproduced with constant shallow heating parameters. This supports the conclusion from \citet{deibel2015} and \citet{parikh2018} for MAXI J0556-332 that not all outbursts of the same source need the same shallow heating. If only one of the shallow heating parameters is tied between outbursts, the $\chi^2$ is close to that of the primary model, but it requires a larger variation of the shallow heating parameter that is not tied between outbursts.

Although the shallow heating parameters of the primary model  are not extreme compared to other sources, one might ask whether or not it is possible that the amount of shallow heating changes between or even during outbursts. For some of the suggested origins of shallow heating it might be possible that the shallow heating is time-variable. If the origin of the shallow heating is related to the nuclear reactions taking place in the crust \citep[e.g. through additional electron captures, fusion reactions, or due to uncertainties in the nuclear symmetry energy, ][]{estrade2011,horowitz2008,steiner2012}, is seems unlikely that the amount of heat released per accreted nucleon would vary per outburst, since those reaction rates should be directly proportional to the accretion rate. Alternatively, if the strong decay in the initial part of the cooling curve is caused by compositionally driven convection in the ocean due to chemical separation of light and heavy elements at the ocean-crust boundary \citep[which reduces the need of additional heat source,][]{horowitz2007,medin2011,medin2014,medin2015}, then the rate of decay should depend on the ocean composition. Although the ocean composition will evolve over time, it is unclear if the required level of variation between outbursts can be obtained over the short timescales of the outbursts of Aql X-1. Additionally, this mechanism cannot account for the large amount of shallow heating required for MAXI J0556-332. 

Another mechanism that has been proposed for the additional heat required to model the observed cooling curves originates from differential rotation between the liquid ocean and the solid crust as matter forms a levitation belt around the equator of neutron star when it accretes from a disk \citep{inogamov2010}. In this model, gravity waves excited in the atmosphere spin up this layer which extends down to the ocean. At the ocean-crust boundary turbulent breaking causes heat release that might account for the shallow heating of the outmost layer of the neutron star crust. The amount of shallow heat than can be released by this mechanism depends, among other things, on how far the accretion disk extends towards the surface of the neutron star. This would suggest that the amount and of shallow heating can be time-variable, because the inner disk radius can vary in time as well \citep[see e.g. the review by][]{done2007}. It would also allow the depth to be time-variable, because the depth of the ocean can change in time.

Finally, the shallow heating strength might depend on the geometry of the accretion flow, which is time-variable. The accretion process can be spherical, for example when the source accretes from a coronal flow or some sort of radiative inefficient accretion flow, or non-spherical when the accretion disk reaches (close to) the neutron star surface. It is often assumed that quasi-spherical accretion is related to the hard spectral state during outburst and non-spherical accretion to the soft state. It might be that the shallow heating process can only be activated in one of the two circumstances or that the shallow heating process is non-uniform when the accretion process is too. 

If indeed the spectral hardness of the emission during outburst is related to the accretion geometry \citep[as e.g.][proposed for accreting black holes]{esin1997}, one would expect that the shallow heating might be related to a specific accretion state. Our models in which we allowed shallow heating to be active only when the source is in either the soft state or the hard state did not provide constraining results. During the full extent of outbursts 14 and 23 the source was observed to be in the hard state, but while outbursts 23 can be modelled without shallow heating (i.e. if shallow heating in only active in the soft state), this seems unlikely for outburst 14. However, it should be emphasised that cooling curve 14 consists of only 2 observations, and therefore we cannot draw strong conclusions based on this cooling curve.

On the other hand, if we activate shallow heating only when the source is in the hard state the amounts of shallow heating per outburst had to be different by a factor $\sim10$ in order to fit all cooling curves. This is due to the fact that during outbursts 20 and 21 the source is in the hard state only for short durations at the start and end of the outbursts, while their cooling curves require a large amount of deposited heat, which can only be accounted for if the shallow heat during these short lived hard state episodes is very strong. If the shallow heating is state-dependent the cooling curves of Aql X-1 can be fitted best if the strength and depth are allowed to change. We have not found any evidence that shallow heating can be constant between outbursts if it is active only during one specific spectral state.

\subsection{Constraining envelope compositions from type-I X-ray bursts}
For the primary model we find that the thickness of the helium layer in the envelope following from the best-fit light element column depths are for all outbursts smaller than the column depth limit for which an X-ray burst is expected, except for outburst 23. The cooling curve of this outburst is well fitted with a column depth that provides a helium layer that is slightly larger than the ignition column depth, but the $1\sigma$ lower limit of the envelope composition lies well below the expected ignition depth. Moreover, to ignite a type-I X-ray burst the temperature in the crust also has to be sufficient, and this is especially in the decay of the outburst not necessarily the case. 

In this paper we tried to constrain the envelope composition not only from the cooling curves, but also from the amount of matter accreted since the last reported observed X-ray bursts during the outburst.  All upper limits on the helium column depths that we calculated in this way (see Table \ref{tab:outburstproperties}) exceed the level required to trigger a burst. Therefore, it seems likely that more bursts have happened after the last reported detected one. If that is indeed the case, at least one other layer of light elements would have been fused into heavier elements and the actual column depth of light elements at the end of the outburst would be lower than we calculated.

Although the calculated upper limits on the thickness of the helium layer in the envelope for Aql X-1 seem unlikely, this method to constrain the envelope composition can be used in future models of crust cooling X-ray bursters. The envelope composition is one of the parameters that are difficult to constrain \citep[unless a significant part of the cooling curve is sampled, see][]{cumming2017}, and have a significant influence on the calculated cooling curve. If the outburst decay is well sampled with observations, this would allow precise determination of the end of the outburst as well as the accretion rate during the decay. On top of those improvements this would increase the chance of detecting the last X-ray burst during the outburst. Alternatively, simulations of such events might allow to determine when the last type-I X-ray burst has occurred during an outburst \citep{johnston2017}. Although further theoretical development is required to accurate determine the times of bursts, this would be a very powerful tool to constrain the envelope composition. If an X-ray burst is detected or theoretically predicted late into de outburst decay, combining this date with exact knowledge of the time-dependent accretion rate during the outburst decay and the outburst time, would allow for a better constrained upper limit of the light element column depth. With better constraints on the envelope composition the shallow heating and core temperature can in consequence be constrained to higher accuracy as well.

\subsection{Importance of determining the correct end of the outburst}\label{sec:discussion-endtime}

Our crust cooling model of Aql X-1 deviates from that of \citet[][]{waterhouse2016}. The main differences are that we model all outbursts collectively rather than individually to take into account any remaining effects from preceding outbursts, as well as that we use a time-dependent accretion rate. Additionally, we recalculated the end dates of the outbursts. For outbursts 21, we determined the end of the outburst to be five days later compared to \citet{waterhouse2016}. As a consequence, one {\it Swift}/XRT observation that was first considered to be obtained when the source was in quiescence is now considered to be taken when the outburst was not fully over yet. An argument that this observation was taken in quiescence is that the source spectrum did not require a power law component. Although the origin of this component is unknown, it is often associated with accretion \citep[see the discussion in][]{wijnands2015}. Therefore, if this observation was taken when the source was still accreting it is unknown why this power-law was not observed. On the other hand, the observation lies significantly above the trend of quiescence XRT observations (see Figure \ref{fig:decay_fit}), favouring the decay interpretation.  

The models that we carried out to investigate the consequences of using the end date determined by \citet{waterhouse2016} and using their accretion assumptions, showed that the steep initial decay that is required with this end date cannot be reproduced unless a constant accretion rate during outburst is assumed (Figure \ref{fig:2012_models}). However, this is inconsistent with the observed light curve. When the accretion rate is determined from the light curve, the crust of the neutron star already starts to cool during the outburst decay, resulting in a shallower cooling curve \citep[see also][]{ootes2016}. Consequently, the potential first cooling observation cannot be fitted along with the next three cooling observations with the model that accurately models the outburst light curve. We checked that the difference between the two models in time-averaged accretion rate over the full outburst in combination with the difference in assumed outburst duration does not influence this outcome.

Although the decay of outburst 21 was relatively well sampled with {\it Swift}/XRT, the sampling rate of once per $1-5$ days is insufficient to constrain the outburst end date to one day. Our findings show that it is critical to have well-sampled monitoring observations during the decay of the outburst to constrain the end time of the outburst. Additionally, from comparison of two models discussed here, we find once more that taking into account accretion rate variability has a significant influence on the calculated cooling curves, in agreement with \citet{ootes2016}.   

\subsection{Accretion flares}\label{discussion:flares}
\citet{cotizelati2014} reported the detection of two accretion flares after outburst 20. We took these flares into account in our model of Aql X-1, and find that when the same model parameters are used for the flares as for outburst 20, the accretion flares have no influence on the calculated cooling curve. The accretion rate during the flares (determined from the light curve) is low enough that the flares do not cause an increase in the temperature during cooling curve 20. Only when the amount of shallow heating during the flares is increased to more than 5.2 MeV nuc$^{-1}$, the $\chi^2$ is affect. This indicates that the observed flares in Aql X-1 do not alter the underlying cooling curve from outburst 20. This is consistent with the conclusion from \citet{fridriksson2011} who calculated potential the effect of flares in XTE J1701 -- 462. 

It should be noted that the presence of an accretion flare can affect the envelope composition. As more matter is dumped on the surface of the neutron star during the flare, the light element column depth increases. However, the calculated average accretion rate during the reported flares in Aql X-1 is so low ($\sim 10^{-4}\text{ }\dot{M}_\text{Edd}$), that the increase in light envelope column depth would be respectively $2.8\times10^{7}\text{ g cm}^{-2}$ and $1.1\times10^{7}\text{ g cm}^{-2}$, which falls within the error bar on the envelope composition of outburst 20 for the primary model. However, if the accretion rate during the reflare is larger, or if the assumed light element column depth at the end of the outburst is smaller, this effect on the envelope composition could be more significant and has to be taken into account. The observed effective temperature will increase after a reflare if the light element column depth increases, but it is also possible that a reflare causes a significant drop in the observed cooling curve if an X-ray burst occurs during the accretion reflare. The light element column depth can in the latter case be smaller after the reflare than after the main outburst.

\section{Conclusions}
By extending our crust cooling code \texttt{NSCool} to model multiple outbursts, we were able to model the accretion history of Aql X-1 from 1996 up to 2015. Assuming that the quiescence observations of this source reported on by \citet{waterhouse2016} can be interpreted as cooling of the crust we fitted the cooling curves to constrain the crustal parameters of the neutron star. Our main conclusions are:

\begin{itemize}
\item Aql X-1 does not have enough time between outbursts to restore crust-core equilibrium. Therefore, the lowest observed temperature of Aql X-1 is expected to be higher than the base level. Additionally, since thermal equilibrium is not restored between outbursts, the calculated cooling curves are strongly dependent on the accretion history. It is therefore important to take previous outbursts into account when modelling transients with relatively short recurrence times.
\item We find that the observed quiescence observation of Aql X-1 are well fitted if the envelope composition and shallow heating parameters (both strength and depth) are allowed to vary between outbursts. If both the shallow heating depth and strength are tied between the outbursts, the data are not well reproduced. Attempts to connect shallow heating during an accretion outburst to one spectral state were unfruitful. 
\item We were not able to constrain the presence of a high impurity pasta layer in Aql X-1, because the quiescence periods between the outbursts are shorter than the thermal timescale of the deepest layers of the crust. 
\item We attempted to constrain the envelope composition from the time of the last observed type-I X-ray burst. Although all upper limits on the column depth of the light element layer exceed the column depth required to ignite a new burst, this method can be used in future studies if a type-I burst is observed shortly before the end of the accretion outburst.
\item The two small accretion reflares observed after one of the outburst were found to have no significant effect on our calculations. 
\end{itemize}



\section*{Acknowledgements}
LSO and RW are supported by a NWO TOP Grant, module 1, awarded to RW. DP is partially supported by the Mexican Conacyt (CB-2014-01, \#240512). ND acknowledges support from an NWO Vidi grant.

This work benefited from discussions at the Physics and Astrophysics of Neutron Star Crusts workshop 2016 supported by the National Science Foundation under Grant No. PHY-1430152 (JINA Center for the Evolution of the Elements).

We made use of the \textit{Swift}/BAT transient monitor results provided by the \textit{Swift}/BAT team, the MAXI data provided by RIKEN, JAXA and the MAXI team. This paper uses preliminary analysis results from the Multi-INstrument Burst ARchive (MINBAR), which is supported under the Australian Academy of Science's Scientific Visits to Europe program, and the Australian Research Council's Discovery Projects and Future Fellowship funding schemes. This research has made use of NASA's Astrophysics Data System Bibliographic Services of the data supplied by the UK Swift Science Data Centre at the University of Leicester. 




\bibliographystyle{mnras}
\bibliography{Aql_X-1} 



\newpage
\onecolumn
\appendix

\begin{figure}
\section{Hard and Soft State}\label{appendix}
	\includegraphics[width=0.9\columnwidth]{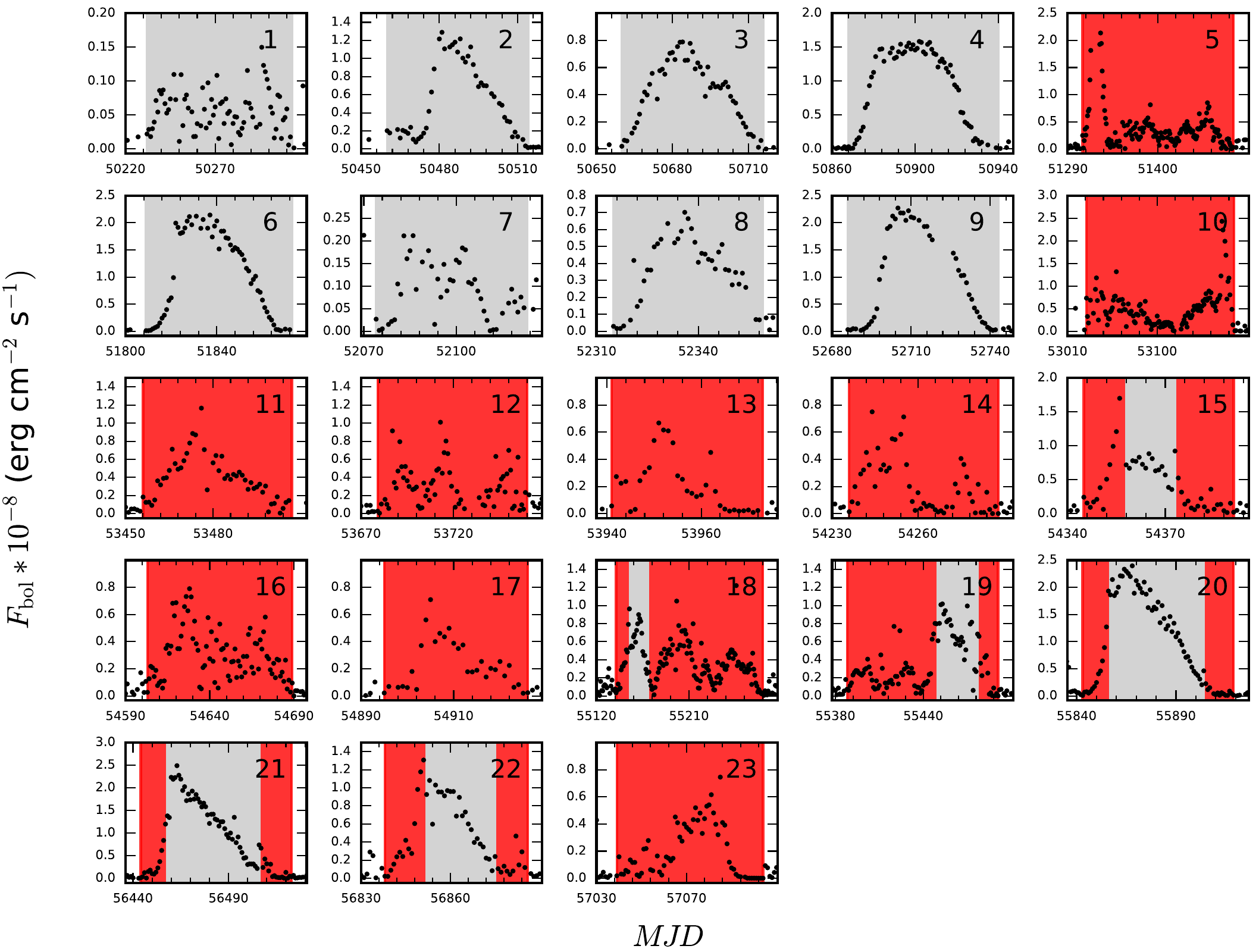}    
	\caption{Light curves of the 23 accretion outbursts of Aql X-1 that are modelled in this work obtained from the combined {\it RXTE}/ASM, {\it Swift}/BAT, and MAXI monitoring observations, and the {\it Swift}/XRT observations. In each frame we indicate when the source was in the soft state (grey areas) and hard state (red areas) during the outburst (observations in white areas are before the start of after the end of the outburst). The spectral state of the source is determined as described in Section \ref{sec:conversion}. For outbursts $1-10$, we determined the spectral state for each full outburst, whereas for outbursts $11-23$ the available data allowed us to distinguish state transitions of the source during an outburst. The number in the upper right corner of each frame indicates the number of the outburst as referred to elsewhere in this paper. }
    \label{fig:Aql_X-1_individual_outbursts_hardsoft}
\end{figure}
\FloatBarrier

\begin{landscape}
\begin{table}
\section{Model parameters}
\setlength{\extrarowheight}{0.3cm}
\centering
\caption{Best-fit parameters with $1\sigma$ errors obtained for the different models described in this work. Since we allow the light element column depth, shallow heating strength and shallow heating depth to vary per outburst, we provide these parameters per outburst horizontally, where the numbering of outbursts is as described in Section \ref{sec:lightcurve}. Outbursts 14, 19, 20, 21, and 23 are the outbursts for which we have quiescence observations and hence are the outbursts for which we can constrain the parameters. "Other" refers either to the 18 outbursts for which we do not have quiescence observations and for which we have fixed the input parameters (indicated by the absence of errors), or to all outbursts (those with and without quiescence observations) in the case that we have tied the parameter between all outbursts. In the latter case the parameter is provided only for "other" and not for individual outbursts, because the value is the same for all outbursts. \newline Model 1 is the primary model, which has the envelope and shallow heating parameters untied for all individual outbursts with quiescence observations. In model 2 the envelope composition is tied between outbursts. In model 3 both shallow heating parameters are tied, while in models 4 only the shallow heating strength and in model 5 the shallow heating depth is tied between outbursts. In model 6 the envelope composition and both shallow heating parameters are tied. Finally, in model 7 the parameters of outbursts 20 and 21 are tied. }
\label{allmodelparams}
\resizebox{1.35\textwidth}{!}{%
\begin{tabular}{lllllllllllllllllllll}
\hline
 Model & $\tilde T_0$                & \multicolumn{6}{l}{$\log{y_\text{L}}$}                                                                                        & \multicolumn{6}{l}{$Q_\text{sh}$}                                                                                                   & \multicolumn{6}{l}{$\rho_\text{sh}$}                                                                                              & $\chi^2$ \\
  & $\times 10^7$ K      & \multicolumn{6}{l}{g cm$^{-2}$}                                                                                                & \multicolumn{6}{l}{MeV nuc$^{-1}$}                                                                                                  & \multicolumn{6}{l}{$\times 10^9$ g cm $^{-3}$}                                                                                    &       \\
  &                      & 14                  & 19               & 20                  & 21                  & 23                  & Other               & 14                  & 19                   & 20                  & 21                  & 23                   & Other               & 14               & 19                   & 20                  & 21                  & 23                    & Other               &          \\ \hline
1 & $\ \ 8.9^{+2.3}_{-1.5}$  & $8.2^{+1.2}_{*}$    & $8.8^{*}_{*}$    & $8.3^{+0.7}_{-0.9}$ & $8.8^{+1.1}_{-1.5}$ & $9.8^{*}_{-2.5}$    & $8.0$               & $2.9^{+1.6}_{-1.2}$ & $1.3^{+1.4}_{-1.0}$  & $3.7^{+1.5}_{-0.9}$ & $2.3^{+0.5}_{-0.3}$ & $\ \ 0.9^{+0.9}_{-0.6}$  & $1.5$               & $0.1^{+1.3}_{*}$ & $\ \ 0.3^{+34.6}_{*}$    & $0.4^{+7.9}_{*}$    & $0.4^{+0.7}_{-0.3}$ & $\ \ 2.8^{+13.8}_{*}$     & $0.4$               & $12.7$    \\
2 & $12.8^{+1.0}_{-5.3}$ &                     &                  &                     &                     &                     & $6.5^{+2.5}_{*}$    & $2.7^{+0.8}_{-0.6}$ & $1.5^{+0.5}_{-0.2}$  & $4.1^{+1.9}_{-1.4}$ & $2.5^{+0.3}_{-0.1}$ & $16.4^{+4.3}_{-3.0}$ & $1.5$               & $0.1^{+0.9}_{*}$ & $\ \ 0.2^{+12.3}_{*}$    & $0.2^{+7.4}_{*}$    & $0.7^{+0.8}_{-0.6}$ & $17.4^{+4.4}_{-2.4}$  & $0.4$               & $12.9$    \\
3 & $\ \ 9.2^{+0.3}_{-0.8}$  & $8.8^{+0.4}_{-0.4}$ & $8.1^{+0.7}_{*}$ & $9.0^{+0.3}_{-0.2}$ & $10.0^{+0.3}_{-0.5}$ & $9.3^{+0.5}_{-0.1}$ & $8.0$               &                     &                      &                     &                     &                      & $1.9^{+0.4}_{-0.1}$ &                  &                      &                     &                     &                       & $0.3^{+0.6}_{*}$    & $27.1$    \\
4 & $\ \ 8.0^{+2.2}_{-0.8}$  & $8.7^{+0.4}_{-0.7}$ & $3.8^{+7.0}_{*}$ & $8.7^{+0.6}_{-1.5}$ & $9.4^{+0.8}_{-0.5}$ & $10.4^{+0.8}_{-2.3}$ & $8.0$               &                     &                      &                     &                     &                      & $2.8^{+0.3}_{-0.3}$ & $0.1^{+0.9}_{*}$ & $\ \ 1.7^{+0.6}_{-0.4}$  & $1.6^{+4.7}_{-1.4}$ & $0.1^{+0.1}_{*}$    & $13.1^{+13.2}_{-5.1}$ & $0.4$               & $12.9$     \\
5 & $\ \ 8.1^{+2.1}_{-0.6}$  & $8.4^{+1.1}_{-1.6}$ & $10.7^{*}_{*}$    & $8.7^{+0.6}_{-0.7}$ & $9.1^{+1.1}_{-0.8}$ & $10.5^{*}_{-2.9}$    & $8.0$               & $3.0^{+2.7}_{-1.2}$ & $0.8^{+1.6}_{-0.3}$  & $3.8^{+0.9}_{-1.0}$ & $2.3^{+0.4}_{-0.4}$ & $\ \ 0.4^{+0.4}_{-0.3}$  & $1.5$               &                  &                      &                     &                     &                       & $0.3^{+0.5}_{*}$    & $12.7$     \\
6 & $13.0^{+0.5}_{-1.4}$ &                     &                  &                     &                     &                     & $6.8^{+0.7}_{-0.3}$ &                     &                      &                     &                     &                      & $2.6^{+0.2}_{-0.3}$ &                  &                      &                     &                     &                       & $2.0^{+0.3}_{-0.7}$ & $40.4$    \\
7 & $\ \ 8.1^{+2.2}_{-0.9}$  & $8.7^{+1.1}_{-1.1}$ & $4.4^{+4.5}_{*}$ & $8.4^{+0.7}_{-0.4}$ & $8.4^{+0.7}_{-0.4}$ & $10.3^{*}_{-1.1}$    & $8.0$               & $3.0^{+1.3}_{-1.3}$ & $11.8^{+2.2}_{-1.1}$ & $2.6^{+0.3}_{-1.5}$ & $2.6^{+0.3}_{-1.5}$ & $\ \ 0.7^{+1.0}_{-0.6}$  & $1.5$               & $0.1^{+1.2}_{*}$ & $11.3^{+5.2}_{-4.8}$ & $0.8^{+0.7}_{-0.6}$ & $0.8^{+0.7}_{-0.6}$ & $\ \ 4.8^{+17.1}_{*}$     & $0.4$               & $13.2$    \\ \vspace{-0.35cm} \\ \hline
\end{tabular}
}
\end{table}
\end{landscape}
\clearpage

\bsp	
\label{lastpage}
\end{document}